\documentclass[pre,amsmath,amssymb,twocolumn,showpacs,superscriptaddress,longbibliography]{revtex4-1}%,longbibliography
\usepackage{amsfonts,amsmath,amssymb,bm}
\usepackage{graphicx,graphics,float}
\usepackage{bm}
\usepackage{amssymb}
\usepackage{colordvi}
\usepackage{graphicx}
\usepackage{color}
\usepackage[colorlinks=true,linkcolor=blue,citecolor=blue,urlcolor=blue]{hyperref}%
\usepackage{hyperref}
\usepackage{comment}
\usepackage{harpoon}

\newcommand{\ov}[1]{\overline{{#1}}}

\begin{document}

\title{Brownian thermal transistors and refrigerators in mesoscopic systems}

\author{Jincheng Lu}%\email{jincheng_lu@qq.com}
\affiliation{School of physical science and technology \&
Collaborative Innovation Center of Suzhou Nano Science and Technology, Soochow University, Suzhou 215006, China.}

\author{Rongqian Wang}%\email{rongqianwang@sina.com}
\affiliation{School of physical science and technology \&
Collaborative Innovation Center of Suzhou Nano Science and Technology, Soochow University, Suzhou 215006, China.}

\author{Chen Wang}\email{wangchenyifang@gmail.com}
\address{Department of Physics, Zhejiang Normal University, Jinhua, Zhejiang 321004, China}

\author{Jian-Hua Jiang}\email{jianhuajiang@suda.edu.cn}
\affiliation{School of physical science and technology \&
Collaborative Innovation Center of Suzhou Nano Science and Technology, Soochow University, Suzhou 215006, China.}

\date{\today}
\begin{abstract}
Fluctuations are significant in mesoscopic systems and of particular importance in understanding quantum transport. Here, we show that fluctuations can be considered as a resource for the operations of open quantum systems as functional devices. We derive the statistics of the thermal transistor amplification factor and the cooling-by-heating refrigerator efficiency under the Gaussian fluctuation framework. Statistical properties of the stochastic thermal transistor and the cooling-by-heating efficiency are revealed in the linear-response regime. We clarify the unique role of inelastic processes on thermal transport in mesoscopic systems. We further show that elastic and inelastic processes lead to different bounds based on the linear transport coefficients by establishing a generic theoretical framework for mesoscopic heat transport, which treats electron and bosonic collective excitations in an equal-footing manner. The underlying physics are illustrated concretely using a double-quantum-dot three-terminal system, though the theory applies to more general systems.
\end{abstract}


\maketitle

\section{Introduction}

Quantum transport of energy and charge are two fundamental phenomena in mesoscopic physics~\cite{yimry1997book,gchen2005book,DubiRMP,JiangCRP,BENENTI20171}.
However, the literature of mesoscopic physics has never treated them equally.
Charge transport has hitherto attracted the major research attention~\cite{ymblanter2000pr,sdatta2005book,jauho2008book,Rafael,DavidPRL},
whereas energy transport is much harder to quantify in both theory and experiments.
With recent dramatic advance of experimental technology~\cite{Roche2015nc,hartmann2015prl,Thier2015,NP,jaliel-exper},
the direct measurements of heat energy and local temperature at nano- and micro-scales become available.
%Charge transport is mainly carried by electrons and holes~\cite{JiangNJP,Tang},
%while energy transport can be carried by electrons~\cite{PED,Jiang2017}, phonons~\cite{OraPRB2010,RenRMP,Yamamoto},
%electron-hole pairs~\cite{Electron-Hole,BijayJiang}, and other neutral collective excitations~\cite{Dubi,wangPRApplied}.
%In particular, thermoelectric transport,  has been intensively explored~\cite{SanchezPRL,SanchezDemon,ArracheaPRL}.
In particular, heat conduction via bosonic excitations considerably fertilizes the family of thermal transport~\cite{gchen2005book,RenRMP,segal2016arpc,Cuieaam6622,ZHANG20201}.
Therefore, theories are anticipated to fully embrace the rich features in heat transport, though some intriguing heat transport effects have been unraveled, e.g., thermal transistors and diodes (i.e., analog of transistors and diodes with charge replaced by heat)~\cite{bli2004prl,bli2006apl,Jiangtransistors,Transistor9,Transistor1} and the cooling-by-heating effect (i.e., cooling of a cold bath by interacting with two hot baths without relying on external forces)~\cite{amari2012prl,bcleuren2012prl,hartle2018prb}.

On the other hand, the rise of mesoscopic thermoelectrics
 has partially revived the study on heat transport in mesoscopic systems~\cite{ewohlman2010prb,rsanchez2011prb,David2011PRB,Jiang2013,Verley2014,JiangNearfield,trade-off}.
The thermoelectric effect,
which enables the conversion of heat into electricity and vice versa, can be exploited to harvest waste heat and covert the heat to useful electric power~\cite{rowe2018thermoelectrics},
Highly efficient thermoelectric devices require salient control over heat and charge conduction~\cite{DiSalvo703,rwhitney2014prl,ewohlman2014pre,CooperativeSpin}.
With the knowledge of mesoscopic physics, it may be possible to harness heat and charge transport in unprecedented ways.
However, heat transport due to bosonic collective excitations is still underestimated, compared with electron  transport in mesoscopic systems.

In this work, we present a generic theoretical framework for mesoscopic heat transport which treats electron and bosonic collective excitations in an equal-footing manner.
This allows us to classify energy transport processes in mesoscopic multiterminal systems as elastic and inelastic transport processes~\cite{ewohlman2010prb,rsanchez2011prb,Jiang2012,sothmann2012prb,Jiangtransistors}.
We show that the fundamental probability and energy conservations are fulfilled in different manners in these two categories, leading to distinct bounds on the linear thermal transport coefficients.
Particularly, in multi-terminal mesoscopic systems, the bounds on the thermal transport Onsager matrix for elastic transport is much stronger than that for the inelastic transport.
Moreover, for three-terminal systems such strong bound forbids thermal transistor effect and cooling-by-heating effect to appear in the linear-transport regime.
While for inelastic transport class, all transport phenomena allowed by the laws of thermodynamics but beyond such a strong bound can be realized,
due to the existence of cooperative energy exchange among multiple reservoirs. We further extend such conclusion to multi-terminal systems where decompositions into hierarchical cooperative transport effects are illustrated. Finally, we reveal the fluctuations of Brownian thermal transistors in the linear-response regime.

\section{Elastic and inelastic processes}
\begin{figure}[htb]
\begin{center}
\centering\includegraphics[width=8.5cm]{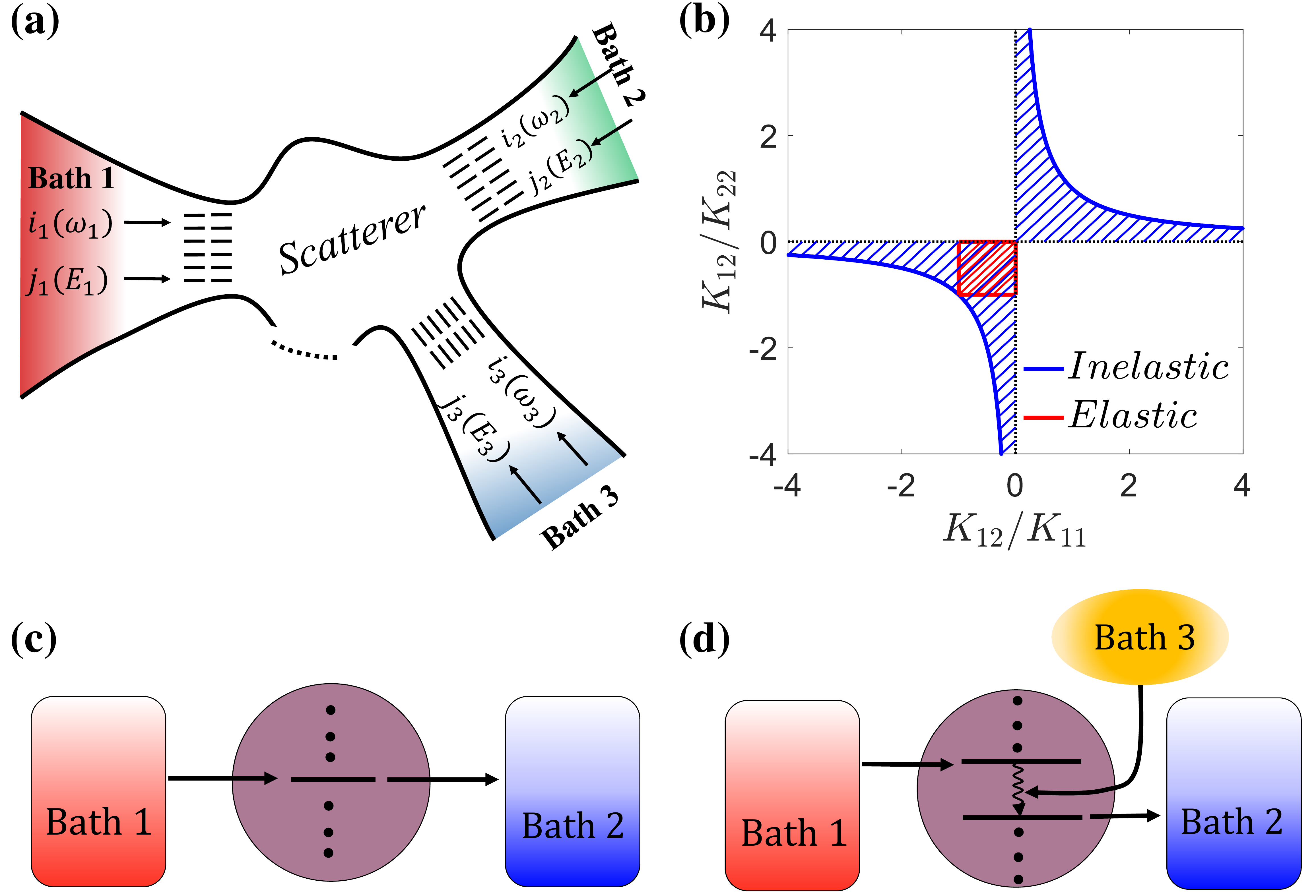}
\caption{(a) Schematic illustration of the multi-terminal energy device. The scattering region is connected to different reservoirs, each of these is able to exchange heat with the system. (b) The boundary of the Onsager coefficients. The shaded blue area represents the broadening of the inelastic case, the shaded red area represents the broadening of the elastic case. (c) Illustration of possible elastic transport processes. (d) Illustration of possible boson-assisted inelastic transport processes.}\label{fig:bound}
\end{center}
\end{figure}

Generally, thermal transport in multi-terminal systems at nanoscale can be modeled in Fig.~\ref{fig:bound}(a), which is driven by thermodynamics bias (e.g., temperature gradient or voltage bias). The steady state transport is characterized by the constant electronic heat currents ($J_m(E)$) and bosonic heat currents ($I_m(\omega)$) which flow into the {\it{m}}th reservoir. During the nonequilibrium exchange processes, there exist two main classes: i) elastic and ii) inelastic transport. And both elastic and inelastic classes contribute to thermal transport~\cite{Jiangtransistors,lu-PRB}.

Specifically, the current densities contributed by elastic (denoted by the subscript `{\it{el}}') processes is expressed by the seminal Landauer's formula~\cite{Sivan,butcher1990}
\begin{subequations}
\begin{align}
&j^{el}_m(E)=\sum_{n\ne m}{\mathcal{T}_{n\rightarrow m}^ef_m(E) - \mathcal{T}_{m\rightarrow n}^ef_n(E)},\\
&i^{el}_m(\omega)=\sum_{n\ne m}{\mathcal{T}_{n\rightarrow m}^bN_m(\omega) - \mathcal{T}_{m\rightarrow n}^bN_n(\omega)}.
\end{align}
\end{subequations}
where $\mathcal{T}_{mn}=|S_{mn}|^2$ is the transmission function with $S_{mn}$ the scattering amplitude via the scatterer in Fig.~\ref{fig:bound}(b),  $f_m=\{\exp[\beta_m(E-\mu_m)]+1\}^{-1}$ is the Fermi-Dirac distribution function with $\beta_m\equiv1/k_BT_m$ the inverse of temperature $T_m$ in the {\it{m}}th fermion bath and $\mu_m$ the corresponding chemical potential, and $N_m=[\exp(\beta_m\omega)-1]^{-1}$ is the Bose-Einstein distribution function in the {\it{n}th} boson bath. Moreover, the probability conservation requires that $\sum_n\mathcal{T}_{n\rightarrow m}^e=\sum_n\mathcal{T}_{m\rightarrow n}^e=1$~\cite{JiangCRP}.
Then, based on the standard Landauer-B\"{u}ttiker theory, the electronic and bosonic heat currents are shown as
\begin{equation}
\begin{aligned}
&J^{el}_m=\frac{1}{h}\int_{-\infty}^{-\infty}dE(E-\mu)j_m^{el(E)},\\
&I^{el}_m=\frac{1}{h}\int_{-\infty}^{-\infty}d\omega\omega i^{el}_m(\omega),
\end{aligned}\label{eq:JmIm}
\end{equation}
respectively. From Eq.~\eqref{eq:JmIm}, it is known that elastic currents (i.e. $J^{el}_m$ and $I^{el}_m$ ) are dominated by two-terminal nonequilibrium processes.

While for the inelastic transport, it is jointly contributed  by multi-terminal processes. The typical realization is the three-terminal nano-device. It is interesting to note that the inelastic current densities into these terminals are the same.
Specifically, inelastic heat currents into three terminals are expressed as by following the Fermi-Golden rule~\cite{Jiang2012,Jiang2013}
\begin{equation}
\begin{aligned}
j_{in}&(E_1,\omega_3) = C_{in}f_1(E_1)[1-f_2(E_1+\omega_3)]N_3(\omega_3)\\
&-C_{in}f_2(E_1+\omega_3)[1-f_1(E_1)][1+N_3(\omega_3)].
\end{aligned}
\end{equation}
where the transition coefficient $C_{in}=\frac{2\pi}{\hbar^2}|v_{in}|^2\rho_{in}(E_1,E_2,\omega_3)\mathcal{T}_1\mathcal{T}_2\mathcal{T}_3$ with $\mathcal{T}_u~(u=1,2,3)$ being the probability for electrons/bosons to tunnel from the $u$th lead into the scatterer, $v_{in}$ is the electron-boson interaction, and $v_{in}$ is the density of states.
Therefore, inelastic heat currents into three terminals are expressed as
\begin{subequations}
\begin{align}
&J^{in}_1=\iint dE_1d\omega_3(E_1-\mu_1)j_{in}(E_1,\omega_3),\\
&J^{in}_2=\iint dE_1d\omega_3(E_1-\mu_2+\omega_3)j_{in}(E_1,\omega_3),\\
&I^{in}_3=\iint dE_1d\omega_3\omega_3j_{in}(E_1,\omega_3),
\end{align}~\label{eq:J1J2I3}
\end{subequations}

We should note that expressions of heat currents for the elastic and inelastic transport at Eq.~\eqref{eq:JmIm} and Eq.~\eqref{eq:J1J2I3} are generally valid both in far-from equilibrium and linear-response regimes. Next, we focus on the thermal transport in linear transport based with the three-terminal setup.

\section{Bound of Onsager coefficients in linear transport}
Typically, for a scatterer interacting with three reservoirs, we have three heat currents, respectively. However, due to the energy conservation ($I_3+\sum_{m=1,2}J_m=0$), we are left with two independent heat currents, e.g., $J_1$ and $I_3$. The transport equation of heat currents can be expressed as~\cite{JiangPRE}
\begin{equation}
\begin{aligned}
\left( \begin{array}{cccc} J_1\\ I_{3} \end{array}\right) =
 \left( \begin{array}{cccc} K_{11} & K_{12} \\ K_{12} &
    K_{22} \end{array} \right) \left( \begin{array}{cccc} \frac{T_1-T_2}{T_2}\\
    \frac{T_3-T_2}{T_2} \end{array}\right),
\end{aligned}
\end{equation}
where $K_{11(22)}$ and $K_{12}$ are the diagonal and off-diagonal
thermal conductance, which was originally derived based
on the Onsager theory, and $|T_{1(3)}-T_2|\ll T_2$. Then, we
describe the bounds of Onsager coefficients in elastic and
inelastic scattering mechanisms separately.

We firstly consider the generic elastic transport.
The elastic coefficients are specified as
\begin{align}
K^{el}_{ij}=\left \langle E^2\right \rangle_{ij} G^{el}_{ij} \quad (i=1,2,3),
\end{align}
where the average under the elastic processes is given by~\cite{MyJAP}
\begin{equation}
\left \langle {\mathcal O(E)} \right \rangle_{ij}=\frac{\int{dE{\mathcal O(E)}G^{el}_{ij}(E)}}{\int{dEG^{el}_{ij}(E)}},
\end{equation}
with the probability weight
\begin{subequations}
\begin{align}
G^{el}_{11}(E)&=({\mathcal T_{12}}+{\mathcal T_{13}})f(E)[1-f(E)],\\
G^{el}_{12}(E)&=(-{\mathcal T_{13}})f(E)[1-f(E)],\\
G^{el}_{22}(E)&=({\mathcal T_{13}}+{\mathcal T_{23}})f(E)[1-f(E)].
\end{align}
\end{subequations}
As the transmission probability $\mathcal T_{ij}{\ge}0$ is positive, it is
straightforward to obtain the boundary of elastic transport coefficients as follows,
\begin{equation}
-1\le{K^{el}_{12}/K^{el}_{22}}\le 0,\quad
-1\le{K^{el}_{12}/K^{el}_{11}}\le 0.
\label{eq:el}
\end{equation}
The above expression is presented graphically in Fig.~\ref{fig:bound}(a) red shadow regime. It is consistent with the results in Ref.~~\cite{Jiangtransistors}, and directly arisen from the second law of thermodynamics.

While for a typical inelastic device consisting of three terminals, the Onsager coefficients are expressed as
\begin{subequations}
\begin{align}
&K^{inel}_{11}=\left \langle E^2_1\right \rangle G^{inel}_{11},\\
&K^{inel}_{12}=\left \langle E_1\omega_3\right \rangle G^{inel}_{12},\\
&K^{inel}_{22}=\left \langle \omega^2_3\right \rangle G^{inel}_{22}.
\end{align}
\end{subequations}
where the ensemble average over all inelastic processes is carried out as
\begin{equation}
\left \langle {\mathcal Q(E,\omega)} \right \rangle=\frac{\iint{dEd\omega{\mathcal Q(E,\omega)}G^{inel}(E,\omega)}}{\iint{dEd{\omega}G^{inel}(E,\omega)}},
\end{equation}
with $G^{inel}=C_{in}f_1(E_1)[1-f_2(E_1+\omega)]N_3(\omega)$. By applying the Cauchy-Schwarz inequality $\left \langle E^2_1\right \rangle\left \langle \omega^2_3\right \rangle-\left \langle E_1\omega_3\right \rangle^2\ge 0$, it is interesting to find that inelastic transport coefficients are bounded by
\begin{equation}
\frac{K^{inel}_{11}}{K^{inel}_{12}}{\times}\frac{K^{inel}_{22}}{K^{inel}_{12}}{\ge}1.
\label{eq:inel}
\end{equation}
We have provided a generic description of linear electronic and bosonic transport in the three-terminal geometry. Remarkably, the two simple relationships Eq.~\eqref{eq:el} and \eqref{eq:inel} hold for all thermodynamic systems in the linear-response regime. The above two relationships bear very important information on the thermoelectric transport, which is one of the main results in the present work. Figure ~\ref{fig:bound}(b) represents them graphically. It is found that the Onsager coefficient for elastic and inelastic classes have dramatically different behaviors which is regardless of the specific mesoscopic systems. Particularly, the inelastic transport coefficient has a much loose bound. In the following, we will show that the inelastic thermal transport in our geometry will realize an thermal transistor, as well as the cooling-by-heating effect.

\section{Bounds of heat amplification}
The quantum thermal transistor~\cite{transistor0,Transistor9,transistor4,transistor-yang,WangPRE} and cooling-by-heating effect~\cite{Cooling1,Cooling2,Cooling3,cooling7} were discovered in thermoelectric devices within linear response regime. It was proposed that the transport mechanism of inelastic scattering plays a crucial role to exhibit the heat amplification, whereas the elastic scattering effect will never show
such amplification behavior~\cite{Jiangtransistors,lu-PRB}.

Recent proposals suggest the proper usage of nonlinearities of a mesoscopic system coupled to environmental modes. A system usually is considered to be connected to two terminals, source (left terminal) and drain (right terminal), and an additional environment boson bath. A temperature distribution $T_j$ ($j=L,R,ph$) generates quantum transport through the system. The aim is to modulate the heat current following out from left terminal $I_L^Q$ with a small modulation of the heat injected from the boson bath $I_{ph}^Q$. This is usually done via inelastic transitions in the system induced by fluctuations in the environment. These can be controlled by tuning the temperature $T_{ph}\rightarrow T_{ph}+\Delta T$ with $\Delta T/T_{ph}\ll1$. Moreover, the thermal transistor can control the heat flow in analogy to the usual electric transistor for the control of the the electric current. A thermal transistor effect appears whose amplification factor is defined as~\cite{Jiangtransistors,Transistor1,lu-PRB}
\begin{equation}
\xi=\frac{\partial_{T_{ph}}I^Q_L}{\partial_{T_{ph}}I^Q_{ph}}.
\label{eq:xi_defination}
\end{equation}
where $I^Q_L$ $(T_{ph}=T)$ is the heat current of the left lead, $I^Q_{ph}$ $(T_{ph}=T)$ is the heat current of the boson bath, and $T$ is the equilibrium temperature of baths. Ref.~~\cite{Jiangtransistors} has proposed a realistic and relatively simple setup for the realization of transistors by exploiting phonon-assisted hopping transport in double quantum dot systems in a three-terminal geometry and found that a thermal transistor effect can
develop in the linear-response regime. Next we will show the bounds of the statistics of stochastic thermal transistor within the three-terminal setup.

We primarily analyze the heat amplification that by modulating the temperature of the boson bath $T_{ph}$, in which the slight change of the boson current may dramatically modulate the current in the left or right electric lead. Following Eq.~\eqref{eq:xi_defination}, the heat current amplification factor can be reexpressed as $\xi=[Q_L^{(1)}-Q_L^{(2)}]/[Q_{ph}^{(1)}-Q_{ph}^{(2)}]$, where $Q_{L(ph)}^{(1)}=I^Q_{L(ph)}(T+\delta T)$ and $Q_{L(ph)}^{(2)}=I^Q_{L(ph)}(T_{ph}=T)$ with $\delta T/T\rightarrow0$. Apart from the stochastic heat current, the average one $\overline{Q}_{L(ph)}^{(i)}$ ($i=1,2$) can characterize our system in the
linear-response regime~\cite{JiangPRE,Jiangtransistors,lu-PRB},
\begin{equation}
\left( \begin{array}{cccc} \overline{Q}_L^{(i)}\\ \overline{Q}_{ph}^{(i)} \end{array}\right) =
 \left( \begin{array}{cccc} K_{11} & K_{12} \\ K_{12} &
    K_{22} \end{array} \right) \left( \begin{array}{cccc} A_L^{(i)}\\
    A_{ph}^{(i)} \end{array}\right),
\end{equation}

Now we introduce the theory of large deviations to analyze the statistics of the currents at long time within the Gaussian approximation~\cite{SeifertPR,PDF1,PDF2,PDF3,Brownian,JiangPRL}. We begin to introduce the probability distribution function of the stochastic heat currents~\cite{Gaspard,Gaspard_2013},
\begin{equation}
\begin{aligned}
P_i(Q_L^{(i)},Q_{ph}^{(i)}) =& \frac{t\sqrt{\det(\hat{K}^{-1})}}{4\pi}\\
&\times\exp\left[-\frac{t}{4}\Delta\vec{Q}_i^T\cdot\hat{K}^{-1}\cdot\Delta\vec{Q}_i\right],
\end{aligned}
\end{equation}
where $\det(\hat{K}^{-1})$ is the determinant of the Onsager response matrix $\hat{K}$  and the superscript '$T$' denotes transpose. While averaged quantities are represented with a bar over the symbols throughout this letter, $\Delta\vec{Q}=\vec{Q}-\vec{\overline{Q}}$ represents fluctuations of the heat currents, where $\vec{\overline{Q}}$ is the average heat current and the is stochastic one. From the probability distribution of stochastic heat currents we calculate the distribution of transistor $P_t(\xi)$ from $h(\xi)=-\lim_{t\rightarrow\infty}\frac{\ln[P_t(\xi)]}{t}$  the large deviation function of stochastic thermal transistor is obtained (see Appendix \ref{app-transistor})
\begin{equation}
h(\xi)=\frac{(K_{12}- K_{22}\xi)^2\Delta A_{ph}^2}{8(K_{11}-2 K_{12}\xi+K_{22}\xi^2)}.
~\label{eq:hxi}
\end{equation}
where we define $\Delta A_{ph}=A_{ph}^{(1)}-A_{ph}^{(2)}$.The large deviation function for three-terminal system, Eq.~\eqref{eq:hxi}, is an another key expression in our work. Its shape can be characterized by the following quantities: average transistor amplification $\overline\xi$ and the width of the distribution around the average, $\sigma_{\xi}$.

%%%%%%%%%%%%%%%%%%%%%%%%%%%%%%%%%%%%%%%%%%%%%%%%%%%%%%%%%%%%%%%%%%
\begin{figure}[htb]
\begin{center}
\centering\includegraphics[width=8.5cm]{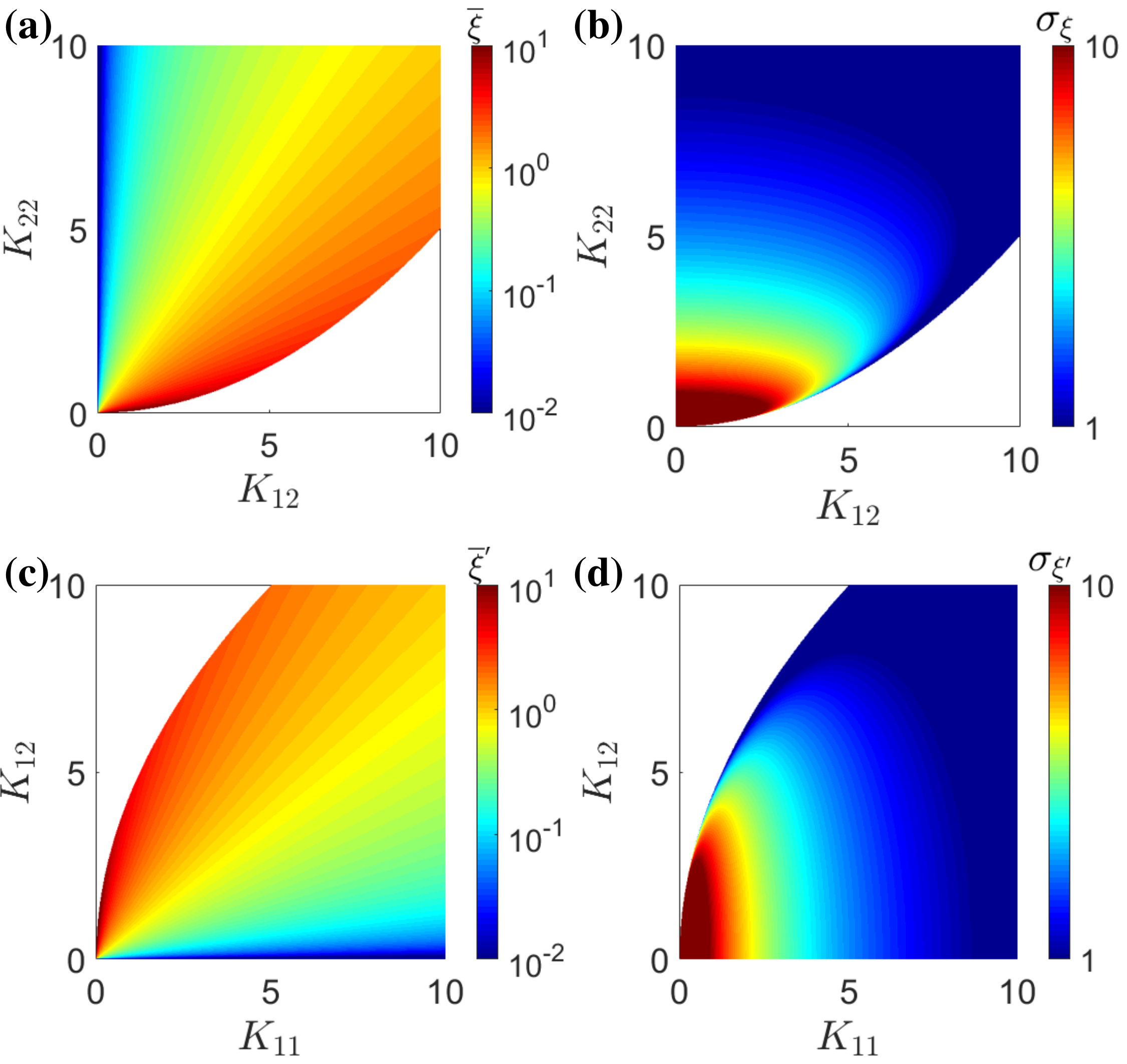}
\caption{(a) $\bar\xi$ and (b) $\sigma_{\xi}$ as functions of $K_{12}$ and $K_{22}$, where $K_{11}=20$, $\Delta A_{ph}=1$. (c) $\bar\xi^{\prime}$ and (d) $\sigma_{\xi^{\prime}}$ as functions of $K_{11}$ and $K_{12}$, where $K_{12}=20$, $\Delta A_L=1$. The white region is forbidden by the thermodynamic bound. } ~\label{fig:transistor}
\end{center}
\end{figure}
%%%%%%%%%%%%%%%%%%%%%%%%%%%%%%%%%%%%%%%%%%%%%%%%%%%%%%%%%%%%%%%%%%

We begin with some general properties of the large deviation function of the stochastic transistor amplification fluctuation. First, $h(\xi)$ has only one minimum and one maximum. Specifically, the minimum $h(\bar\xi)=0$ locates at the average transistor amplification~\cite{Jiangtransistors}
\begin{equation}~\label{ampeff1}
\overline\xi=\frac{K_{12}}{K_{22}},
\end{equation}
which correspond to the maximal probability for the appearance of the amplification efficiency at Eq.~\eqref{ampeff1}. It should  be noted that $\overline\xi$ only relies on the general expression of the transport coefficients $K_{12}$ and $K_{22}$, see Fig.~\ref{fig:transistor}(a). Specifically, for the elastic thermal transport, $\overline{\xi}_{el}$ is always below the unit as $-1<\frac{K_{12}^{el}}{K_{22}^{el}}<0$. While for the inelastic case with the constraint coefficients bound at Eq.~\eqref{eq:inel}, the average efficiency is given by $\overline{\xi}_{in}<\left| \frac{K_{11}^{in}}{K_{12}^{in}}\right|$, which can be modulated in the regime $\overline{\xi}_{in}$. Hence, the stochastic transistor may work as $\frac{K_{11}^{in}}{K_{12}^{in}}>1$.  Moreover, for the inelastic transport case, the Onsager coefficients are constraint by
the second law of thermodynamics, $K_{11}K_{22}-K_{12}^2\ge0$. Therefore, the the bound of amplification average efficiency is given by $0<\overline\xi<\infty$ (blue shadow regime in Fig.~\ref{fig:bound}(b)).

The width of the distribution around the average transistor amplification factor $\overline{\xi}$, which is another key characteristic of transistor amplification fluctuations. Specifically, expanding $h(\xi)$ around its average $\overline{\xi}$, $h(\xi)\simeq \frac{1}{2\sigma_{\xi}^2}(\xi - \overline{\xi})^2 + {\mathcal O}((\xi-\overline{\xi})^3)$, the amplification fluctuation is obtained as
\begin{equation}
\sigma_{\xi}=\frac{2\sqrt{K_{22}(K_{11}K_{22}-K_{12}^2)}}{K_{22}^2\Delta A_{ph}},
\end{equation}
which obeys the bound of the Onsager coefficients $K_{11}K_{22}-K_{12}^2\ge0$ and $K_{22}{\ge}0$~\cite{JiangPRE}. The equality is reached as the fluctuation width completely vanishes. Obviously, when this equality is reached, the total entropy production rate of the system in the linear-response regime $\frac{dS}{dt}\equiv0$, i.e., the system is in the equilibrium state~\cite{PED}. The width $\sigma_{\xi}$ is plotted at Fig.~\ref{fig:transistor}(b) where the white color regime is forbidden according to the second law of thermodynamics. By fixing the off-diagonal coefficient $K_{12}$, $\sigma_{\xi}$ is found to be small when $K_{22}$ is large ($\sigma_{\xi}{\approx}2\sqrt{K_{11}}/(K_{22}{\Delta}A_{ph}$)), which corresponds to the low amplification average efficiency $\overline\xi{\ll}1$. While the fluctuation becomes strong as $K_{22}$ is tuned down, by approaching to the thermodynamic reversible bound.

The three terminals are equivalent for the modulation of the thermal transistor. In addition to the boson bath discussed above, as a comparison, we also study that the thermal transistor is optimally manipulated by the left electronic lead temperature $T_L$, i.e., $\Delta A_{ph}=0$. The heat current amplification factor is similarly defined by
\begin{equation}
\xi'=\frac{\partial_{T_L}I^Q_L}{\partial_{T_L}I^Q_{ph}}.
\end{equation}
And we can obtain the large deviation function of stochastic thermal transistor $h(\xi')=\frac{(K_{12}-\xi' K_{11})^2\Delta A_{L}^2}{8(K_{22}-2\xi' K_{12}+\xi'^2K_{11})}$, where $\Delta A_L=A_L^{(1)}-A_L^{(2)}$. As $\xi'\rightarrow\infty$, we have $h(\xi'\rightarrow\infty)=\frac{1}{8}K_{11}\Delta A_{L}^2$, which corresponds to Eq.~(\ref{eq:hxi}).
For the system under $T_L$ modulation, the average transistor amplification factor is given by
\begin{equation}
\overline\xi'=\frac{K_{12}}{K_{11}},
\label{eq:xibar2}
\end{equation}
and the amplification fluctuation is
\begin{equation}
\sigma_{\xi^{\prime}}=\frac{2\sqrt{K_{11}(K_{11}K_{22}-K_{12}^2)}}{K_{11}^2\Delta A_{L}}.
\end{equation}
It is interesting to find the relationship between Eqs.~\eqref{ampeff1} and \eqref{eq:xibar2},
\begin{eqnarray}
\overline\xi_{in}{\times}\overline\xi^{\prime}_{in}=1,
\end{eqnarray}
at the thermodynamic bound $K_{11}K_{22}-K_{12}^2=0$. Figure \ref{fig:transistor}(c) and \ref{fig:transistor}(d) demonstrate the thermal transistor modulated by the source temperature $T_{L}$ behaviors: the average amplification efficiency $\overline\xi'$ only relies on the general expression of the transport coefficients $K_{11}$ and $K_{12}$, and it reaches its maximum when $K_{11}$ tends to zero. The width of the distribution $\sigma_{\xi^{\prime}}$ is plotted in Fig.~\ref{fig:transistor}(d) where the white region is forbidden by the second law of thermodynamics according to Eq.~\eqref{eq:inel}. It is small when $K_{12}$ is small and $K_{11}$ is large, corresponding to the average amplification efficiency $\overline\xi'$. Approaching the inelastic transport boundary $K_{11}K_{22}=K_{12}^2$, we find that $\sigma_{\xi^{\prime}}\rightarrow \infty$ for $K_{11}\rightarrow0$.

From the above discussion, we can find that the three-terminal system can be operated as an excellent thermal transistor no matter which heat bath is manipulated. The control of mesosocopic fluctuations allows for huge heat amplification factor in the both cases. Controllability makes the operation easily exportable to different kinds of systems and interactions.

\section{Cooling-by-heating effect}
%%%%%%%%%%%%%%%%%%%%%%%%%%%%%%%%%%%%%%%%%%%%%%%%%%%%%%%%%%%%%%%%%%
\begin{figure}[htb]
\begin{center}
\centering \includegraphics[width=8.5cm]{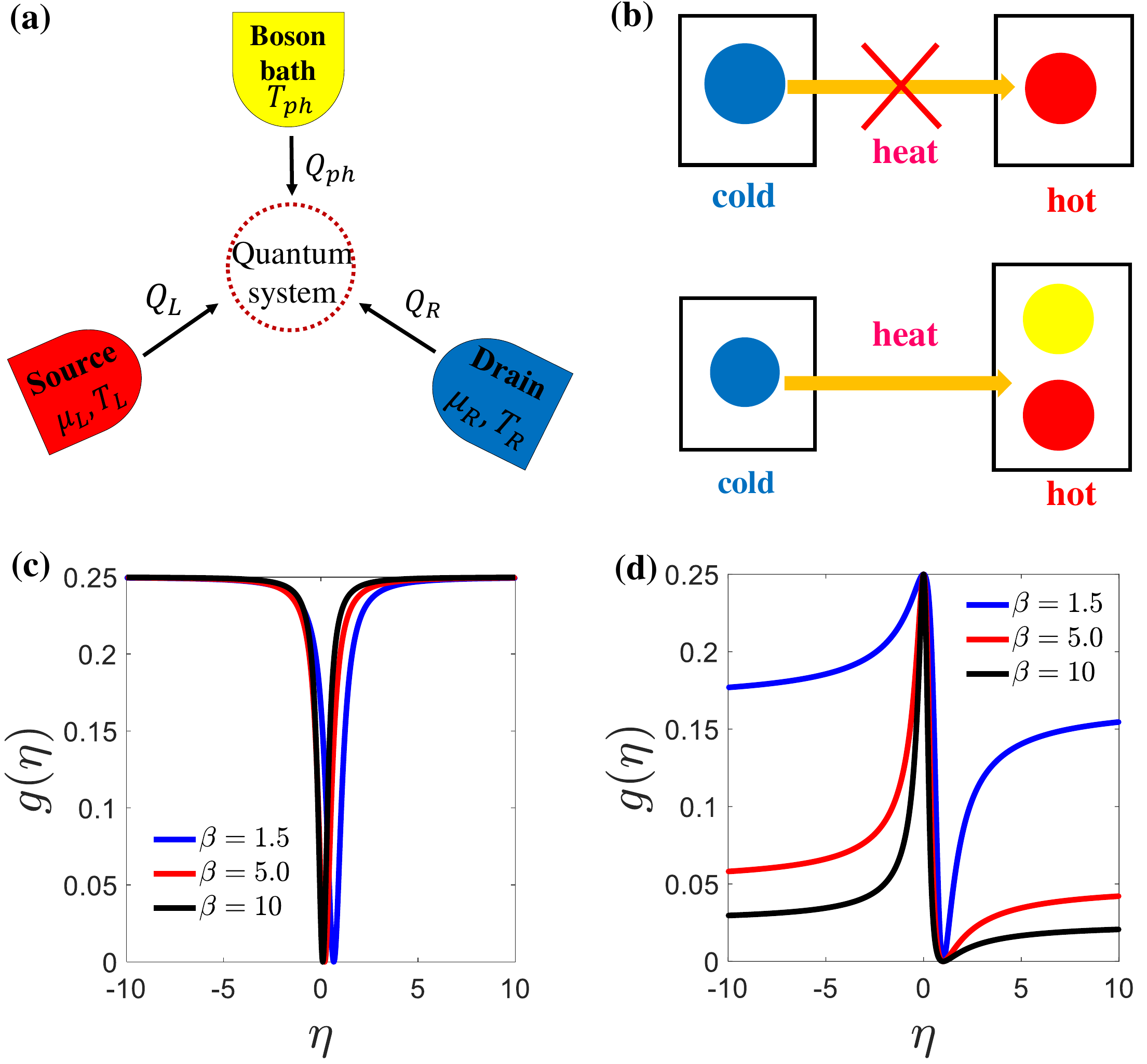}
\caption{(a) Schematic illustration of the three-terminal energy device. The quantum system is connected to two electronic reservoirs (source and drain) and boson bath.  The temperatures and chemical potentials of two electric reservoirs are $\mu_{L(R)}$ and $T_{L(R)}$, respectively. The temperature of boson bath is $T_{ph}$. $Q_i$ ($i=L,R,ph$) represents the heat current following into the quantum system. (b) The up-panel: in a two-terminal system the second law of thermodynamics forbids the heat flow from the cold terminal to the hot one. The down-panel: a cold reservoir is cooled by two hotter ones, i.e., cooling-by-heating. This process is allowed by the second law of thermodynamics and hence it can happen without changing the external world. (c) $g(\eta)$ as a function of $\eta$ for different $\beta$, where $\alpha=1.0$, $d=0$. (d) $g(\eta)$ as a function of $\eta$ for different $\beta$, where $\alpha=1.0$, $d=\infty$. }~\label{fig:cooling}
\end{center}
\end{figure}
%%%%%%%%%%%%%%%%%%%%%%%%%%%%%%%%%%%%%%%%%%%%%%%%%%%%%%%%%%%%%%%%%%

%%{\it{`cooling by heating` effect.}}
This section is devoted to the manifestation of the inelastic process and the second law of thermodynamics in the three-terminal systems. Usually the second law is expressed in a two-terminal fashion. For example, Clausius's statement: ``{\it No process is possible whose sole result is the transfer of heat from a body of lower temperature to a body of higher temperature"}. For three-terminal systems the second law of thermodynamics has a more complex face where some counterintuitive effects can be allowed. For example, in the cooling-by-heating effect where two hot reservoirs can cool a cold one without changing the rest of the world~\cite{Cooling1,Cooling2}. The cooling-by-heating effect in quantum systems can be understood that as the quantum device is driven by the external work, the heat is extracted from the cool bath and absorbed by the hot bath [see Fig.~\ref{fig:cooling}(b)]. Recently, the cooling-by-heating are analyzed in photovoltaic, optomechanic systems, which exploit the refrigeration by photons~\cite{Cooling1,Cooling2}.

Here, we show that the three-terminal thermoelectric system can also be tuned to be a refrigerator~\cite{JiangBijayPRB17,Rongqian,David-refrigerator,Friedman2018,Rafael-Cooling,cooling4}. As shown in Fig.~\ref{fig:cooling}(a), under the influence of the refrigerator, the left reservoir (source) may be cooled. The cooling efficiency is defined as $\eta=\frac{\dot{Q}}{\dot{W}}=\frac{Q_L}{Q_{ph}}$. In the linear-response regime, the scaled large deviation function of stochastic efficiency~\cite{FT6,FT3,Niedenzu,FT8} is (see Appendix \ref{app-cooling})
\begin{equation}
g(\eta)=\frac{[1-\beta\eta+(\alpha-\eta)d]^2}{4[\alpha+\eta(-2+\beta\eta)](\alpha d^2+2d+\beta)},
~\label{eq:geta}
\end{equation}
with dimensionless parameters
\begin{equation}
\alpha=\frac{K_{11}}{K_{12}},\quad
\beta=\frac{K_{22}}{K_{12}},\quad
d=\frac{A_L}{A_{ph}},
\end{equation}
and the thermodynamic forces are $A_L=(T_R-T_L)/T$ and $A_{ph}=(T_{ph}-T_L)/T$, respectively.

In Fig.~\ref{fig:cooling}(c) and \ref{fig:cooling}(d) we illustrate the rich behavior of $g(\eta)$ at various conditions: (i) $d=0$, (ii) $d=\infty$. Note that the $g(\eta)$ is always bounded 0 and $\frac{1}{4}$.
When $d=0$ (i.e., $T_L=T_R$), the large deviation function $g(\eta)$ experiences a sharp transition for $\eta\rightarrow0$ and it behaves like a derivative of the Dirac delta function. The upper bound of $g(\eta)$ is reached at the $|\eta|>2$. In contrast, we find that only when $\eta\approx0$ the $g(\eta)$ reached $\frac{1}{4}$ for $d=\infty$ (i.e., $T_L=T_{ph}$). Moreover, when the efficiency $|\eta|>2$, the large deviation function of stochastic efficiency tends to steady.

%%%%%%%%%%%%%%%%%%%%%%%%%%%%%%%%%%%%%%%%%%%%%%%%%%%%%%%%%%%%%%%%%%
\begin{figure}[htb]
\begin{center}
\centering\includegraphics[width=8.8cm]{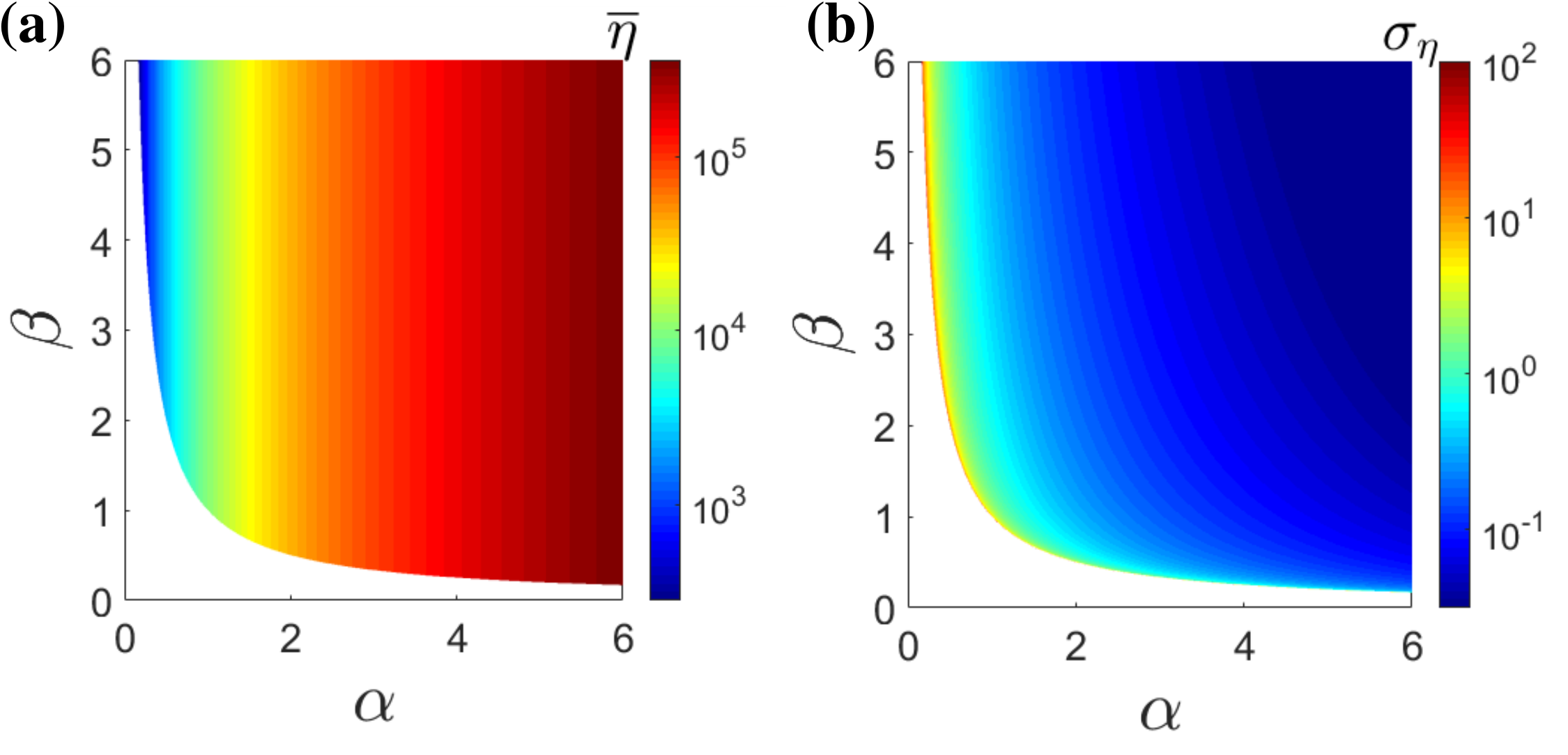}
\caption{(a) $\bar\eta$ and (b) $\sigma_\eta$ as functions of $\alpha$ and $\beta$, where $d=\infty$. The white region is forbidden by the thermodynamic bound.} \label{fig:etabar}
\end{center}
\end{figure}
%%%%%%%%%%%%%%%%%%%%%%%%%%%%%%%%%%%%%%%%%%%%%%%%%%%%%%%%%%%%%%%%%%

The minimum of $g(\bar\eta)=0$ is reached at the average efficiency
\begin{equation}
\bar\eta=\frac{\alpha d+1}{d+\beta}.
\end{equation}
The working regime of the refrigerator can be obtained as $(\alpha d+1)(d+\beta)>0$, as shown in Fig.~\ref{fig:etabar}(a). Specifically, as $T_L=T_R$ ($d=0$) the average efficiency is simplified as $\overline{\eta}_1=K_{12}/K_{22}$. While as $T_L=T_{ph}$ ($d=\infty$), the average efficiency becomes $\overline{\eta}_2=K_{11}/K_{12}$. Then, we discuss the behavior of cooling efficiency in elastic and inelastic, separately. For the elastic thermal transport, it is known that the Onsager coefficients are bounded as $-1<K_{12}/K_{22(11)}<0$. Hence, there is no cooling-by-heating effect in the elastic transport ($\overline{\eta}_{1(2)}<0$).  While for the inelastic thermal transport, the bound of the coefficients is given by $K^2_{12}/K_{11}K_{22}<1$. It is interesting to find that cooling efficiencies are restricted by $\overline{\eta}_1\times\overline{\eta}_2<1$ Moreover, both two cooling efficiencies range from 0 to $\infty$.

The fluctuating width of the average efficiency, $\sigma_\eta$, is another key characteristic of cooling efficiency. By expanding $h(\overline\eta)=0$ around its minimum $\overline\eta$, we obtain
\begin{equation}
\sigma_\eta=\frac{(d+\beta)^2}{(\alpha d^2+2d+\beta)\sqrt{2(\alpha\beta-1)}}.
\end{equation}
Figs.~\ref{fig:etabar}(b) illustrates the behavior of the width of cooling efficiency distribution $\sigma_\eta$ the condition of $d=\infty$. Specifically, as $d=0$ the width is reduced to $\sigma_\eta=\frac{\beta}{\sqrt{2(\alpha\beta-1)}}$. While as $d=\infty$, it is given by $\sigma_\eta=\frac{1}{\sqrt{2(\alpha\beta-1)}}$.
We see in this figure that the $\sigma_{\eta}$ reaches the maximum under the limit condition, $(\alpha d+1)(d+\beta)=0$.

We conclude this section emphasizing central observations: the statistics of cooling efficiency can reveal information on the three-terminal thermoelectric system, and the average efficiency and its fluctuation reach the optimal values at the limit of the second law of thermodynamics and further characterize the properties of the system, as expected.

%%%%%%%%%%%%%%%%%%%%%%%%%%%%%%%%%%%%%%%%%%%%%%%%%%%%%%%%%%%%%%%%%%
\begin{figure}[htb]
\begin{center}
\centering\includegraphics[width=8.5cm]{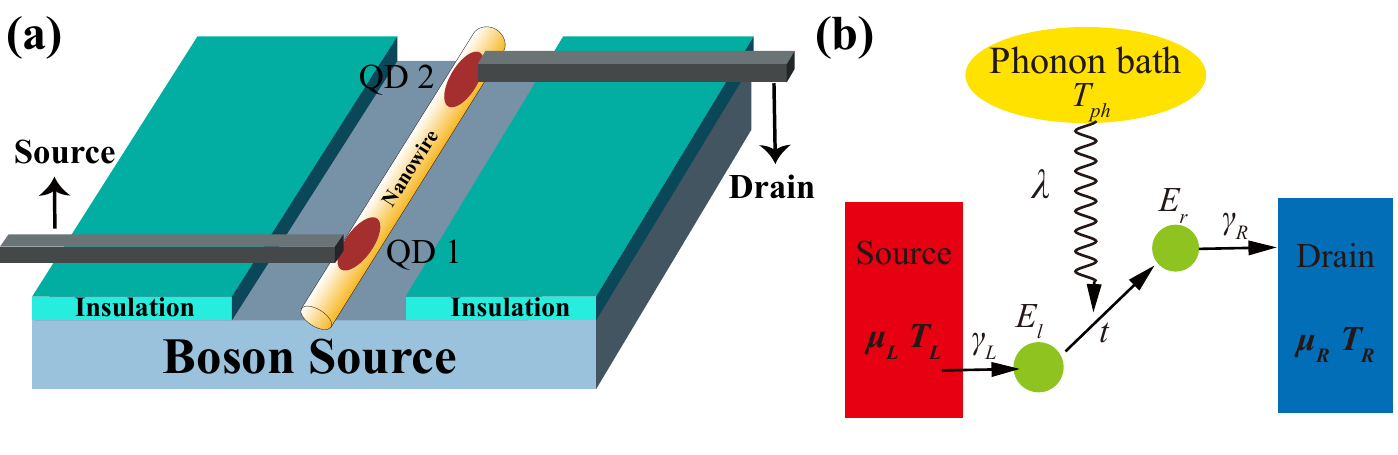}
\caption{(a) Scheme of a double-quantum-dot device that can serve as a thermal transistor. The QDs are embedded in the nanowire and are controlled by gate voltages: $l$ and $r$ control the local potentials, and $t$ tunes the tunneling between the QDs. The two electrodes, $L$ and $R$, apply voltage and temperature biases across the QDs. The insulation layer suppresses the thermal contact between the metal electrodes and the substrate, which provides thermal energy to phonons. (b) Illustration of the three-terminal inelastic transport. An electron left the source into the first QD (with energy $E_l$) hops to the second QD (with a different energy $E_r$) assisted by a phonon from the phonon bath (with temperature $T_{ph}$). Tunneling rates between the dots and the electron leads ($\gamma_L$ and $\gamma_R$) and in between the dots ($t$) can be tuned via gate-controlled tunnel barriers. The electron then tunnels into the drain electrode from the second QD. Such a process gives inelastic charge transfer from the source to the drain assisted by the phonon from the phonon source. Both the process and its time reversal contribute to the inelastic thermoelectricity in the system. The electrochemical potential and temperature of the source (drain) are $\mu_L$ and $T_L$ ($\mu_R$ and $T_R$), respectively.}~\label{fig:DQD}
\end{center}
\end{figure}
%%%%%%%%%%%%%%%%%%%%%%%%%%%%%%%%%%%%%%%%%%%%%%%%%%%%%%%%%%%%%%%%%%

\section{Inelastic thermoelectric transistor in three-terminal double-quantum-dot system}
We exemplify our analysis within a mesoscopic double quantum dots (QDs) thermoelectric device under the time-reversal symmetry. A typical inelastic thermoelectric device consisting of three terminals: two electrodes (the source and the drain) and a boson bath (e.g., a phonon bath). The device is schematically depicted in Fig.~\ref{fig:DQD}(a) and explained in the caption.  The phonon-assisted hopping inelastic transport is illustrated in Fig.~\ref{fig:DQD}(b) and explained in the caption. In phonon-assisted hopping transport, the figure of merit is limited by the average frequency and bandwidth of the phonons (rather than electrons) involved in the inelastic transport~~\cite{Jiangtransistors}.

Specifically, the system is described by the Hamiltonian
\begin{equation}
\hat H = \hat H_{\rm DQD} + \hat H_{\rm e-ph} + \hat H_{\rm lead} + \hat H_{\rm tun} + \hat H_{\rm ph},
\end{equation}
with
\begin{subequations}
\begin{align}
\hat H_{\rm DQD} &= \sum_{i=\ell,r} E_i \hat c_i^\dagger \hat c_i + ( t \hat c_l^\dagger \hat c_r + {\rm H.c.}) , \\
\hat H_{\rm e-ph} &= \lambda \hat c_l^\dagger \hat c_r (\hat a + \hat a^\dagger) + {\rm H.c.} ,\\
\hat H_{\rm ph} &= \omega_{0}\hat a^\dagger \hat a, \\
\hat H_{\rm lead} &= \sum_{j=L, R}\sum_{k} \varepsilon_{j,k} \hat c_{j,k}^\dagger \hat c_{j,k} ,\\
\hat H_{\rm tun} &= \sum_k V_{L, k} \hat c_\ell^\dagger \hat c_{L, k} + \sum_k V_{R, k} \hat c_r^\dagger \hat c_{R, k} +  {\rm H.c.},
\end{align}
\end{subequations}
where $\hat c_i^\dagger$ ($i=\ell, r$) creates an electron in the $i$-th QD with an energy $E_{i}$,
$\lambda$ is the strength of electron-phonon interaction, and
$\hat a^\dagger$($\hat a$) creates(annihilates) one phonon with the frequency $\omega_{0}$.

This noninteracting model has been analyzed thoroughly in Ref.~~\cite{Jiang2012}, to study transport coefficients and the average efficiency. The inelastic contribution to the currents is calculated based on the Fermi golden rule as~\cite{Jiangtransistors}
\begin{equation}
I^Q_L = E_l I_N , \quad I^Q_{ph} = (E_r-E_l) I_N.
\label{current}
\end{equation}
where $I^Q_L = E_l I_N $ and $I^Q_{ph} = (E_r-E_l) I_N$ are the heat currents following from the source and phonon bath, respectively. The particle current
$I_N = \Gamma_{l\rightarrow r} - \Gamma_{r\rightarrow l}$ with $\Gamma_{l \rightarrow r} \equiv \gamma_{e-ph} f_{l} ( 1 - f_r) N_p^-$ and $\Gamma_{r\rightarrow l} \equiv \gamma_{e-ph} f_r (1-f_l) N_p^+$. Here $N_p^{\pm}=N_B+\frac{1}{2}\pm\frac{1}{2}{\rm sgn}(E_r-E_l)$
with the Bose-Einstein distribution $N_B\equiv[\exp(\frac{|E_r-E_l|}{T_{ph}})-1]^{-1}$. The transition
rate $\gamma_{e-ph}= \xi_0 \left(\frac{|E_r-E_l|}{\omega_{0}}\right)^{n} \exp\left[-\left(\frac{E_r-E_l}{\omega_{0}}\right)^2\right]$. Here $\xi_0$ stands for the electron-phonon scattering strength, $n$ provides the power-law dependence on phonon energy $E_r-E_l$ with a characteristic energy $\omega_0$. We assume that the contact between the source and the QD 1 and the contact between the drain and the QD2 can be made very good. Under such conditions, we can approximate the distributions on the QD1 and QD2 can be approximated as $f_l\approx [\exp(\frac{E_l-\mu_L}{k_BT_L})+1]^{-1}$ and $f_r\approx [\exp(\frac{E_r-\mu_R}{k_BT_R})+1]^{-1}$.

%%%%%%%%%%%%%%%%%%%%%%%%%%%%%%%%%%%%%%%%%%%%%%%%%%%%%%%%%%%%%%%%%%
\begin{figure}[htb]
\begin{center}
\centering\includegraphics[width=8.5cm]{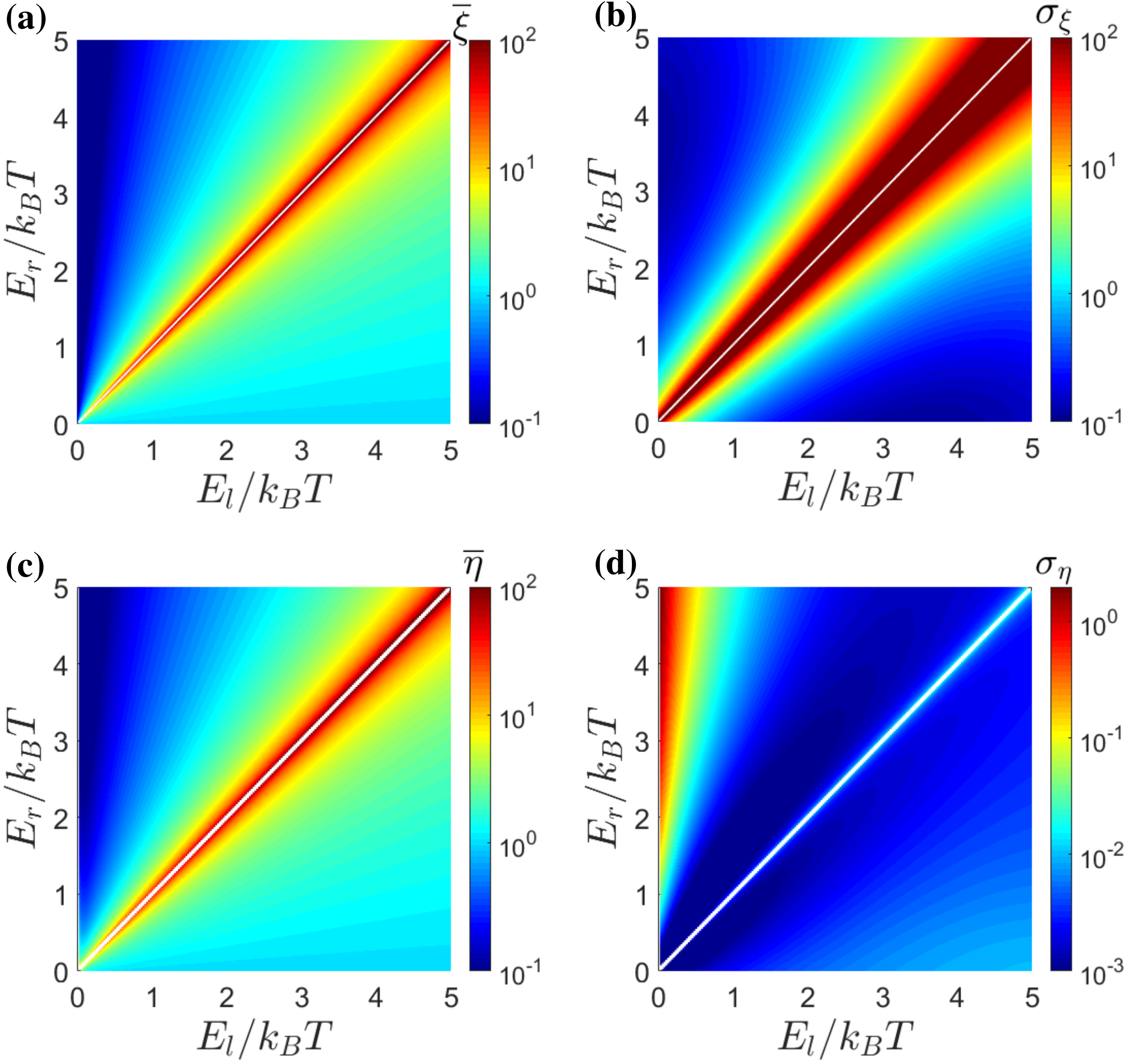}
\caption{(a) The average amplification efficiency $\overline\xi$, (b) width of transistor amplification factor distribution $\sigma_{\xi}$, (c) the average efficiency $\overline\eta$, (d) the fluctuating width of the average efficiency $\sigma_{\eta}$ as the functions of QD energies $E_l$ and $E_r$. The parameters are $\xi_0=0.1k_BT$, $\omega_0=10k_BT$, $n=1$, $\gamma_l=\gamma_r=0.02k_BT$, $\mu_L=\mu_R=0$, $t=0.01k_BT$ and $d=\infty$.}~\label{fig:DQD2}
\end{center}
\end{figure}
%%%%%%%%%%%%%%%%%%%%%%%%%%%%%%%%%%%%%%%%%%%%%%%%%%%%%%%%%%%%%%%%%%

With such a double QD device, the phenomenological Onsager transport equation is written in the linear-response regime as
\begin{equation}
\left( \begin{array}{cccc} I^Q_L \\ I^Q_{ph} \end{array}\right) =
 \left( \begin{array}{cccc} K_{11} & K_{12} \\ K_{12} &
    K_{22} \end{array} \right) \left( \begin{array}{cccc} \frac{T_L-T_R}{T}\\
    \frac{T_{ph}-T_R}{T} \end{array}\right),
\end{equation}
where $K_{11}=\frac{\partial I^Q_L}{\partial T_L}$, $K_{12}=\frac{\partial I^Q_L}{\partial T_{ph}}$, and $K_{22}=\frac{\partial I^Q_{ph}}{\partial T_{ph}}$ in the limit $T_L,T_R,T_{ph}\rightarrow T$.

It is interesting to find that a thermal transistor effect can be realized in the {\em linear-response regime}~\cite{Jiangtransistors}.
The average amplification efficiency of the heat current can be obtained as
\begin{align}
\overline\xi \equiv \left |\frac{\partial_{T_{ph}} I^Q_L}{\partial_{T_{ph}}
    I^Q_{ph}} \right| = \left | \frac{K_{12}}{K_{22}} \right | = \left | \frac{E_l}{E_r-E_l} \right |,
\label{alpha}
\end{align}
which is clearly shown at Fig.~\ref{fig:DQD2}(a). Remarkably, average amplification efficiency $\overline\xi>1$ is achieved with $|E_l|>|E_r-E_l|$, which is solely contributed by the inelastic contribution.
As shown in Fig.~\ref{fig:DQD2}(b), for the amplification fluctuation $\sigma_{\xi}$, the divergent behavior is observed near the regime $E_l{\approx}E_r$, whereas it is strongly suppressed at large energy bias $|E_r-E_l|$. Moreover, if we consider the complete contribution of the inelastic electron-phonon scattering (e.g., $t=0$) to the amplification fluctuation, it is intriguing to obtain the general condition as $\sigma_{\xi}\equiv 0$, regardless of the quantum dot energy levels or electron-phonon coupling strength.

The cooing by heating effect also can be realized in this three-terminal double QDs system. The average cooling efficiency $\overline\eta$ and the fluctuating width of the average efficiency $\sigma_{\eta}$ as shown in Figs.~\ref{fig:DQD2}(c) and \ref{fig:DQD2}(d).
We find that $\overline\eta$ is significantly enhanced around the regime $E_l\approx E_r$, whereas $\sigma_\eta$ is dramatically suppressed.
The maximum of fluctuating width $\sigma_{\eta}$ is obtained when $E_l<1.0k_BT$ and $E_r>2.0k_BT$.
Hence, the analysis statistical influence of the fluctuation  on the effectiveness of cooing by heating effect can be straightforwardly conducted in the optimal system parameter regimes.

\section{Conclusion and discussions}
In this work, we demonstrate that fluctuations can be exploited to enable two kinds of anomalous heat transport phenomena: the linear thermal transistor effect and the cooling-by-heating effect. Normally, thermal transistor effects take place in the nonlinear transport regime and negative differential thermal conductance is required. In Brownian thermal transistors, thermal transistor effects can appear without such conditions, as powered by accidental fluctuations. Similarly, the condition for the cooling-by-heating effect is loosed. Nevertheless, the statistics reveals more information. We derive the statistical distributions of the thermal amplification factor and the cooling-by-heating efficiency under the Gaussian fluctuation framework. Two main statistical properties, namely the average value and the variance, of the thermal amplification factor and the cooling-by-heating efficiency, are studied and discussed to uncover more information of the mesoscopic transport.

We further reveal the unique role of inelastic processes in thermal transport in mesoscopic systems. We show that elastic and inelastic transport processes lead to distinct bounds on the linear transport coefficients by establishing a generic theoretical framework for mesoscopic heat transport which treats electron and bosonic collective excitations in an equal-footing manner. With such knowledge, we demonstrate that based on the ensemble-averaged performances, the linear thermal transistor effect and the cooling-by-heating effect can take place only for thermal transport based on inelastic processes. The underlying physics is illustrated concretely using a double-quantum-dot three-terminal system, though the theory applies to more general systems.
The theoretical framework established in this work paves the way for the study of coupled thermal and electrical transport and fluctuations in generic multiterminal mesoscopic systems, which is on the horizon for both theory and experiments. Finally, it should be pointed out that our study is based on the linear-response theory with time-reversal symmetry. Future generalizations to nonlinear and time-reversal broken regime are anticipated and demanded.

\section{Acknowledgment}
J.L., R.W., and J.-H.J. acknowledge support from the National Natural Science Foundation of China (NSFC Grant No. 11675116), the Jiangsu distinguished professor funding and a Project Funded by the Priority Academic Program Development of Jiangsu Higher Education Institutions (PAPD). C.W. is supported by the National Natural Science Foundation of China under Grant No. 11704093. %We gratefully acknowledge useful discussions with Manas Kulkarni and Jie Ren.

\appendix

\section{Derivation of the large deviation function for thermal transistor effect}~\label{app-transistor}

We begin by introducing the probability distribution function (PDF) of the stochastic heat currents $Q_{L(ph)}^{(i)}$, ($i=1,2$)
\begin{equation}
P_i(Q_L^{(i)},Q_{ph}^{(i)})=\frac{t\sqrt{\det(\hat{K}^{-1})}}{4\pi}\exp\left[-\frac{t}{4}\Delta\vec{Q}_i^T\cdot\hat{K}^{-1}\cdot\Delta\vec{Q}_i\right]
\end{equation}
where $\det(\hat{K}^{-1})$ is the determinant of the symmetric part of the inverse of the Onsager response matrix $\hat{K}$ and the superscript $"T"$ denotes transpose. While averaged quantities are represented with a bar over the symbols throughout this Letter, $\delta\vec{Q}=\vec{Q}-\vec{\overline{Q}}$ represents fluctuations of the heat currents, where $\overline{Q}_L^{(i)}$ and $\overline{Q}_{ph}^{(i)}$ are the average heat current and photonic current, respectively, the ${Q}_L^{(i)}$ and ${Q}_{ph}^{(i)}$ are stochastic ones.

From the probability distribution of stochastic heat currents we calculate the distribution of thermal transistor $P_t(\xi)$
\begin{equation}
\begin{aligned}
P_t(\xi)& = \int dQ_L^{(1)}dQ_L^{(2)}dQ_{ph}^{(1)}dQ_{ph}^{(2)}P_1(Q_L^{(1)},Q_{ph}^{(1)})\\
&\times P_1(Q_L^{(2)},Q_{ph}^{(2)})\delta\left(\xi-\frac{Q_L^{(1)}-Q_L^{(2)}}{Q_{ph}^{(1)}-Q_{ph}^{(2)}}\right).
\end{aligned}
\end{equation}
Replacing the stochastic heat currents with fluctuations of the heat currents
\begin{equation}
{Q}_L^{(i)}-\overline{Q}_L^{(i)}=\Delta {Q}_L^{(i)}, \quad {Q}_{ph}^{(i)}-\overline{Q}_{ph}^{(i)}=\Delta {Q}_{ph}^{(i)}.
\end{equation}
So now we can rewrite the PDF of stochastic heat current as
\begin{equation}
\begin{aligned}
P_t(\xi)&=\int d(\Delta Q_L^{(1)})d(\Delta Q_L^{(2)})d(\Delta Q_{ph}^{(1)})d(\Delta Q_{ph}^{(2)})\\
&\times P_1(\Delta Q_L^{(1)}\Delta Q_{ph}^{(1)})P_2(\Delta Q_L^{(2)},\Delta Q_{ph}^{(2)})\\
&\times \delta\left(\xi-\frac{Q_{L}^{(1)}-Q_{L}^{(2)}+\ov{Q}_{L}^{(1)}-\overline{Q}_{L}^{(2)}}{Q_{ph}^{(1)}-Q_{ph}^{(2)}+\overline{Q}_{ph}^{(1)}-\ov{Q}_{ph}^{(2)}}\right).
\end{aligned}
\end{equation}
A direct calculation yields an expression
\begin{equation}
P_t(\xi)=C\exp\left[\frac{[(K_{11}-\xi K_{12})\Delta A_{ph}+(K_{12}-\xi K_{22})\Delta A_L]^2}{8(K_{11}-2\xi K_{12}+\xi^2K_{22})}\right],
\end{equation}
where we have defined $\Delta A_L=A_L^{(1)}-A_L^{(2)}$ and $\Delta A_{ph}=A_{ph}^{(1)}-A_{ph}^{(2)}$, $C$ is a complex constant. Finally, the large deviation function of stochastic thermal transistor is obtained as
\begin{equation}
\begin{aligned}
h(\xi)&=-\lim_{t\to\infty}\frac{\ln[P_t(\xi)]}{t}\\
&=\frac{[(K_{11}-\xi K_{12})\Delta A_{ph}+(K_{12}-\xi K_{22})\Delta A_L]^2}{8(K_{11}-2\xi K_{12}+\xi^2K_{22})}.
\end{aligned}
\end{equation}

%%%%%%%%%%%%%%%%%%%%%%%%%%%%%%%%%%%%%%%%%%%%%%%%%%%%%%%%%%%%%%%%%%
\begin{figure}[htb]
\begin{center}
\centering\includegraphics[width=8.5cm]{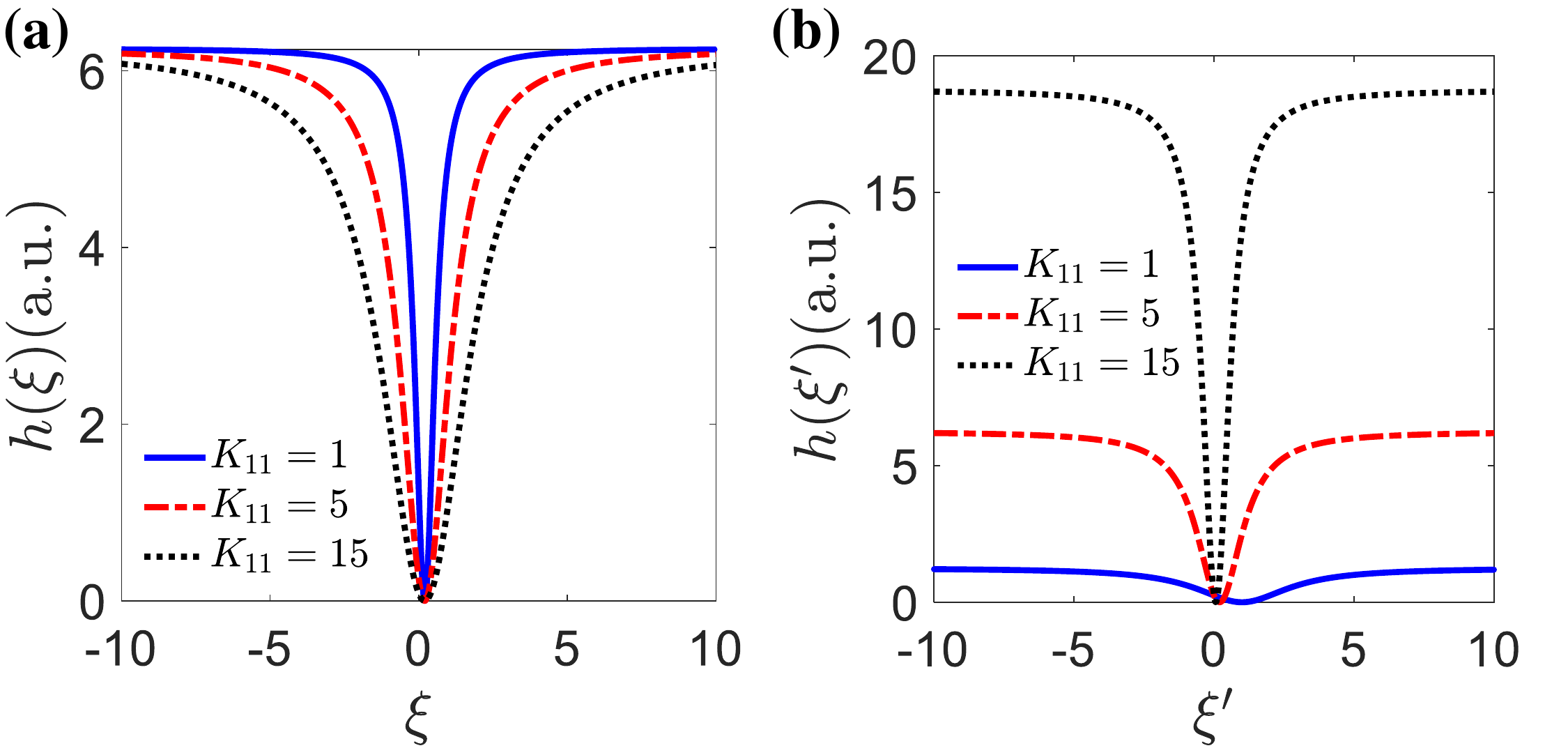}
\caption{(a) $h(\xi)$ as a function of $\xi$ for different $K_{11}$ where $K_{12}=1$, $K_{11}=5$ and $\Delta A_{ph}=10^{-3}$. (b) $h(\xi^{\prime})$ as a function of $\xi^{\prime}$ for different $K_{11}$ where $K_{12}=1$, $K_{11}=5$ and $\Delta A_{L}=10^{-3}$.}\label{fig:hxi}
\end{center}
\end{figure}
%%%%%%%%%%%%%%%%%%%%%%%%%%%%%%%%%%%%%%%%%%%%%%%%%%%%%%%%%%%%%%%%%%

We examine the efficiency statistic of Fig.~\ref{fig:hxi}(a) and \ref{fig:hxi}(b) at different $K_{11}$. We find that in the limit $\xi\rightarrow0$, the large deviation function experiences a sharp transition. Therefore, both $h(\xi)$ and $h(\xi^{\prime})$ behaves like a Lorentz function. When increasing the value of $K_{11}$, the broadening of $h(\xi)$ and $h(\xi^{\prime})$ grows, and as expected, the $h(\xi)$ and $h(\xi^{\prime})$ becomes increasing unreliable. Moreover, when the transistor amplification $\xi$ tends to be large, the deviation function $h(\xi)$ and $h(\xi^{\prime})$ gradually tends to be same.

\section{Derivation of the large deviation function for cooling-by-heating effect}\label{app-cooling}

The three-terminal device can be tuned to be a refrigerator, by exchanging temperatures of the electrode and the boson bath, i.e., $T_c=T_L$ and $T_R=T_h$, with $T_h>T_c$. Then, the left reservoir can be cooled, and heat $Q_L$ is transferred to the boson reservoir. This is cooling-by-heating effect.

In the linear-response regime, the transport equations are expressed as
\begin{equation}
\begin{aligned}
\left( \begin{array}{cccc} \ov{Q}_L\\ \ov{Q}_{ph} \end{array}\right) =
 \left( \begin{array}{cccc} K_{11} & K_{12} \\ K_{12} &
    K_{22} \end{array} \right) \left( \begin{array}{cccc} A_L\\
    A_{ph} \end{array}\right),
\end{aligned}
\end{equation}
where $A_L=(T_h-T_c)/T$ and $A_{ph}=(T_{ph}-T_c)/T$. $\ov{Q}_{L(ph)}$ represents average currents. The efficiency of refrigerator is defined as  $\eta=\frac{\dot{Q}}{\dot{W}}=\frac{Q_L}{Q_{ph}}$.

We begin by introducing the probability distribution function (PDF) of the stochastic heat currents~\cite{JiangPRL}
\begin{equation}
P_t(Q_L,Q_{ph})=\frac{t\sqrt{\det(\hat{K}^{-1})}}{4\pi}\exp\left(-\frac{t}{4}\Delta\vec{Q}^T\cdot\hat{K}^{-1}\cdot\Delta\vec{Q}\right),
\end{equation}
where $\Delta\vec{Q}=\vec{Q}-\vec{\bar Q}$, $\vec{Q}^T=(Q_L,Q_{ph})^T$ is the fluctuations of heat current, $\vec{\bar Q}^T=(\bar Q_L,\bar Q_{ph})^T$ and $\det(\hat{K}^{-1})$ is the determinant of matrix $\hat{K}$. The PDF of stochastic efficiency is
\begin{equation}
\begin{aligned}
P_t(\eta)&=\int_{-\infty}^{\infty} dQ_LdQ_{ph}P_t(Q_L,Q_{ph})\delta\left(\eta-\frac{Q_L}{Q_{ph}}\right)\\
&=\int_{-\infty}^{\infty} dQ_{ph}|Q_{ph}|P_t(\eta Q_{ph},Q_{ph}).
\end{aligned}
\end{equation}
After calculating we get
\begin{equation}
\begin{aligned}
P_t(\eta Q_{ph},Q_{ph})&=\frac{t\sqrt{\det(\hat{K}^{-1})}}{4\pi}\\
&\times\exp\left[-\frac{t}{4}[a(\eta)Q_{ph}^2+2b(\eta)Q_{ph}+c]\right]
\end{aligned}
\end{equation}
with
\begin{subequations}
\begin{align}
&a(\eta)=\frac{(K_{11}-2K_{12}\eta+M_{22}\eta^2)}{\det(\hat{K})},\\
&b(\eta)=\frac{[{K_{12}\bar{Q}_L-K_{11}\bar{Q}_{ph}-\eta(K_{22}\bar{Q}_L-K_{12}\bar{Q}_{ph})}]}{\det(\hat{K})},\\
&c=\frac{(K_{22}\bar{Q}^2_{L}-2K_{12}\bar{Q}_L\bar{Q}_{ph}+K_{11}\bar{Q}^2_{ph})}{\det(\hat{K})}.
\end{align}
\end{subequations}
The full probability distribution of the stochastic efficiency is now found to be
\begin{equation}
\begin{aligned}
P_t(\eta) &= \frac{\sqrt{\det(\hat{K}^{-1})}\exp(-t/4)}{2\pi a(\eta)\sqrt{a(\eta)}}\times
[b\sqrt{\pi t}\exp\left(\frac{b^2(\eta)t}{4a(\eta)}\right) \\
&+2\sqrt{a(\eta)}-b\sqrt{\pi t}\exp\left(\frac{b^2(\eta)t}{4a(\eta)}\right){\rm erf}\left(\frac{b(\eta)\sqrt{t}}{2\sqrt{a(\eta)}}\right)].
\end{aligned}
\end{equation}
The large deviation function of stochastic efficiency is obtained from  $g(\eta)=-\lim_{t\rightarrow\infty}\frac{\ln[P_t(\eta)]}{t\bar{S}_{tot}}$
\begin{equation}
\begin{aligned}
g(\eta) = \frac{[(K_{11}-K_{12}\eta)A_L+(K_{12}-K_{22}\eta)A_{ph}]^2}{4\bar{S}_{tot}[K_{11}+\eta(-2K_{12}+K_{22}\eta)]},
\end{aligned}
\end{equation}
where $\bar{S}_{tot}=\ov{Q}_LA_L+\ov{Q}_{ph}A_{ph}$. By substituting the following parametrization,
\begin{equation}
\alpha=\frac{K_{11}}{K_{12}}, \ \beta=\frac{K_{22}}{K_{12}}, \ d=\frac{A_L}{A_{ph}}.
\end{equation}
we find that
\begin{equation}
g(\eta)=\frac{[(\alpha-\eta)d+(1-\beta\eta)]^2}{4[\alpha+\eta(-2+\beta\eta)](\alpha d^2+2d+\beta)}.
\end{equation}

%\begin{equation}
%\begin{aligned}
%g(\eta) &= \frac{[(K_{11}-K_{12}\eta)A_L+(K_{12}-K_{22}\eta)A_{ph}]^2}{4\bar{S}_{tot}[K_{11}+\eta(-2K_{12}+K_{22}\eta)]}\\
%&\times \frac{1}{(K_{11}A_L^2+2K_{12}A_LA_{ph}+K_{22}A_{ph}^2)},
%\end{aligned}
%\end{equation}

\bibliography{Ref-transistor}

%merlin.mbs apsrev4-1.bst 2010-07-25 4.21a (PWD, AO, DPC) hacked
%Control: key (0)
%Control: author (0) dotless jnrlst
%Control: editor formatted (1) identically to author
%Control: production of article title (0) allowed
%Control: page (1) range
%Control: year (0) verbatim
%Control: production of eprint (0) enabled
\begin{thebibliography}{73}%
\makeatletter
\providecommand \@ifxundefined [1]{%
 \@ifx{#1\undefined}
}%
\providecommand \@ifnum [1]{%
 \ifnum #1\expandafter \@firstoftwo
 \else \expandafter \@secondoftwo
 \fi
}%
\providecommand \@ifx [1]{%
 \ifx #1\expandafter \@firstoftwo
 \else \expandafter \@secondoftwo
 \fi
}%
\providecommand \natexlab [1]{#1}%
\providecommand \enquote  [1]{``#1''}%
\providecommand \bibnamefont  [1]{#1}%
\providecommand \bibfnamefont [1]{#1}%
\providecommand \citenamefont [1]{#1}%
\providecommand \href@noop [0]{\@secondoftwo}%
\providecommand \href [0]{\begingroup \@sanitize@url \@href}%
\providecommand \@href[1]{\@@startlink{#1}\@@href}%
\providecommand \@@href[1]{\endgroup#1\@@endlink}%
\providecommand \@sanitize@url [0]{\catcode `\\12\catcode `\$12\catcode
  `\&12\catcode `\#12\catcode `\^12\catcode `\_12\catcode `\%12\relax}%
\providecommand \@@startlink[1]{}%
\providecommand \@@endlink[0]{}%
\providecommand \url  [0]{\begingroup\@sanitize@url \@url }%
\providecommand \@url [1]{\endgroup\@href {#1}{\urlprefix }}%
\providecommand \urlprefix  [0]{URL }%
\providecommand \Eprint [0]{\href }%
\providecommand \doibase [0]{http://dx.doi.org/}%
\providecommand \selectlanguage [0]{\@gobble}%
\providecommand \bibinfo  [0]{\@secondoftwo}%
\providecommand \bibfield  [0]{\@secondoftwo}%
\providecommand \translation [1]{[#1]}%
\providecommand \BibitemOpen [0]{}%
\providecommand \bibitemStop [0]{}%
\providecommand \bibitemNoStop [0]{.\EOS\space}%
\providecommand \EOS [0]{\spacefactor3000\relax}%
\providecommand \BibitemShut  [1]{\csname bibitem#1\endcsname}%
\let\auto@bib@innerbib\@empty
%</preamble>
\bibitem [{\citenamefont {Imry}(1997)}]{yimry1997book}%
  \BibitemOpen
  \bibfield  {author} {\bibinfo {author} {\bibfnamefont {Y.}~\bibnamefont
  {Imry}},\ }\href@noop {} {\emph {\bibinfo {title} {Introduction to Mesoscopic
  Physics}}}\ (\bibinfo  {publisher} {Oxford University Press, London},\
  \bibinfo {year} {1997})\BibitemShut {NoStop}%
\bibitem [{\citenamefont {Chen}(2005)}]{gchen2005book}%
  \BibitemOpen
  \bibfield  {author} {\bibinfo {author} {\bibfnamefont {G.}~\bibnamefont
  {Chen}},\ }\href@noop {} {\emph {\bibinfo {title} {Nanoscale Energy Transport
  and Conversion}}}\ (\bibinfo  {publisher} {Oxford University Press, London},\
  \bibinfo {year} {2005})\BibitemShut {NoStop}%
\bibitem [{\citenamefont {Dubi}\ and\ \citenamefont
  {Di~Ventra}(2011)}]{DubiRMP}%
  \BibitemOpen
  \bibfield  {author} {\bibinfo {author} {\bibfnamefont {Y.}~\bibnamefont
  {Dubi}}\ and\ \bibinfo {author} {\bibfnamefont {M.}~\bibnamefont
  {Di~Ventra}},\ }\bibfield  {title} {\enquote {\bibinfo {title} {Heat flow and
  thermoelectricity in atomic and molecular junctions},}\ }\href {\doibase
  10.1103/RevModPhys.83.131} {\bibfield  {journal} {\bibinfo  {journal} {Rev.
  Mod. Phys.}\ }\textbf {\bibinfo {volume} {83}},\ \bibinfo {pages} {131}
  (\bibinfo {year} {2011})}\BibitemShut {NoStop}%
\bibitem [{\citenamefont {Jiang}\ and\ \citenamefont {Imry}(2016)}]{JiangCRP}%
  \BibitemOpen
  \bibfield  {author} {\bibinfo {author} {\bibfnamefont {J.-H.}\ \bibnamefont
  {Jiang}}\ and\ \bibinfo {author} {\bibfnamefont {Y.}~\bibnamefont {Imry}},\
  }\bibfield  {title} {\enquote {\bibinfo {title} {Linear and nonlinear
  mesoscopic thermoelectric transport with coupling with heat baths},}\ }\href
  {\doibase https://doi.org/10.1016/j.crhy.2016.08.006} {\bibfield  {journal}
  {\bibinfo  {journal} {C. R. Phys.}\ }\textbf {\bibinfo {volume} {17}},\
  \bibinfo {pages} {1047} (\bibinfo {year} {2016})}\BibitemShut {NoStop}%
\bibitem [{\citenamefont {Benenti}\ \emph {et~al.}(2017)\citenamefont
  {Benenti}, \citenamefont {Casati}, \citenamefont {Saito},\ and\ \citenamefont
  {Whitney}}]{BENENTI20171}%
  \BibitemOpen
  \bibfield  {author} {\bibinfo {author} {\bibfnamefont {G.}~\bibnamefont
  {Benenti}}, \bibinfo {author} {\bibfnamefont {G.}~\bibnamefont {Casati}},
  \bibinfo {author} {\bibfnamefont {K.}~\bibnamefont {Saito}}, \ and\ \bibinfo
  {author} {\bibfnamefont {R.~S.}\ \bibnamefont {Whitney}},\ }\bibfield
  {title} {\enquote {\bibinfo {title} {Fundamental aspects of steady-state
  conversion of heat to work at the nanoscale},}\ }\href {\doibase
  10.1016/j.physrep.2017.05.008} {\bibfield  {journal} {\bibinfo  {journal}
  {Phys. Rep.}\ }\textbf {\bibinfo {volume} {694}},\ \bibinfo {pages} {1}
  (\bibinfo {year} {2017})}\BibitemShut {NoStop}%
\bibitem [{\citenamefont {Blanter}\ and\ \citenamefont
  {B\"{u}ttiker}(2000)}]{ymblanter2000pr}%
  \BibitemOpen
  \bibfield  {author} {\bibinfo {author} {\bibfnamefont {Y.~M.}\ \bibnamefont
  {Blanter}}\ and\ \bibinfo {author} {\bibfnamefont {M.}~\bibnamefont
  {B\"{u}ttiker}},\ }\bibfield  {title} {\enquote {\bibinfo {title} {Shot noise
  in mesoscopic conductors},}\ }\href {\doibase 10.1016/S0370-1573(99)00123-4}
  {\bibfield  {journal} {\bibinfo  {journal} {Phys. Rep.}\ }\textbf {\bibinfo
  {volume} {336}},\ \bibinfo {pages} {1} (\bibinfo {year} {2000})}\BibitemShut
  {NoStop}%
\bibitem [{\citenamefont {Datta}(2005)}]{sdatta2005book}%
  \BibitemOpen
  \bibfield  {author} {\bibinfo {author} {\bibfnamefont {S.}~\bibnamefont
  {Datta}},\ }\href@noop {} {\emph {\bibinfo {title} {Quantum Transport: Atom
  to Transistor}}}\ (\bibinfo  {publisher} {Cambridge University Press,
  London},\ \bibinfo {year} {2005})\BibitemShut {NoStop}%
\bibitem [{\citenamefont {Haug}\ and\ \citenamefont
  {Jauho}(2008)}]{jauho2008book}%
  \BibitemOpen
  \bibfield  {author} {\bibinfo {author} {\bibfnamefont {H.}~\bibnamefont
  {Haug}}\ and\ \bibinfo {author} {\bibfnamefont {A.~P.}\ \bibnamefont
  {Jauho}},\ }\href@noop {} {\emph {\bibinfo {title} {Quantum Kinetics in
  Transport and Optics of Semiconductors}}}\ (\bibinfo  {publisher}
  {Springer-Verlag Berlin Heidelberg},\ \bibinfo {year} {2008})\BibitemShut
  {NoStop}%
\bibitem [{\citenamefont {S\'anchez}\ and\ \citenamefont
  {B\"uttiker}(2011{\natexlab{a}})}]{Rafael}%
  \BibitemOpen
  \bibfield  {author} {\bibinfo {author} {\bibfnamefont {R.}~\bibnamefont
  {S\'anchez}}\ and\ \bibinfo {author} {\bibfnamefont {M.}~\bibnamefont
  {B\"uttiker}},\ }\bibfield  {title} {\enquote {\bibinfo {title} {Optimal
  energy quanta to current conversion},}\ }\href {\doibase
  10.1103/PhysRevB.83.085428} {\bibfield  {journal} {\bibinfo  {journal} {Phys.
  Rev. B}\ }\textbf {\bibinfo {volume} {83}},\ \bibinfo {pages} {085428}
  (\bibinfo {year} {2011}{\natexlab{a}})}\BibitemShut {NoStop}%
\bibitem [{\citenamefont {S\'anchez}\ and\ \citenamefont
  {L\'opez}(2013)}]{DavidPRL}%
  \BibitemOpen
  \bibfield  {author} {\bibinfo {author} {\bibfnamefont {D.}~\bibnamefont
  {S\'anchez}}\ and\ \bibinfo {author} {\bibfnamefont {R.}~\bibnamefont
  {L\'opez}},\ }\bibfield  {title} {\enquote {\bibinfo {title} {Scattering
  theory of nonlinear thermoelectric transport},}\ }\href {\doibase
  10.1103/PhysRevLett.110.026804} {\bibfield  {journal} {\bibinfo  {journal}
  {Phys. Rev. Lett.}\ }\textbf {\bibinfo {volume} {110}},\ \bibinfo {pages}
  {026804} (\bibinfo {year} {2013})}\BibitemShut {NoStop}%
\bibitem [{\citenamefont {Roche}\ \emph {et~al.}(2015)\citenamefont {Roche},
  \citenamefont {Roulleau}, \citenamefont {Jullien}, \citenamefont {Jompol},
  \citenamefont {Farrer}, \citenamefont {Ritchie},\ and\ \citenamefont
  {Glattli}}]{Roche2015nc}%
  \BibitemOpen
  \bibfield  {author} {\bibinfo {author} {\bibfnamefont {B.}~\bibnamefont
  {Roche}}, \bibinfo {author} {\bibfnamefont {P.}~\bibnamefont {Roulleau}},
  \bibinfo {author} {\bibfnamefont {T.}~\bibnamefont {Jullien}}, \bibinfo
  {author} {\bibfnamefont {Y.}~\bibnamefont {Jompol}}, \bibinfo {author}
  {\bibfnamefont {I.}~\bibnamefont {Farrer}}, \bibinfo {author} {\bibfnamefont
  {D.A.}\ \bibnamefont {Ritchie}}, \ and\ \bibinfo {author} {\bibfnamefont
  {D.C.}\ \bibnamefont {Glattli}},\ }\bibfield  {title} {\enquote {\bibinfo
  {title} {Harvesting dissipated energy with a mesoscopic ratchet},}\ }\href
  {\doibase 10.1038/ncomms7738} {\bibfield  {journal} {\bibinfo  {journal}
  {Nat. Comm.}\ }\textbf {\bibinfo {volume} {6}},\ \bibinfo {pages} {6738}
  (\bibinfo {year} {2015})}\BibitemShut {NoStop}%
\bibitem [{\citenamefont {Hartmann}\ \emph {et~al.}(2015)\citenamefont
  {Hartmann}, \citenamefont {Pfeffer}, \citenamefont {H\"ofling}, \citenamefont
  {Kamp},\ and\ \citenamefont {Worschech}}]{hartmann2015prl}%
  \BibitemOpen
  \bibfield  {author} {\bibinfo {author} {\bibfnamefont {F.}~\bibnamefont
  {Hartmann}}, \bibinfo {author} {\bibfnamefont {P.}~\bibnamefont {Pfeffer}},
  \bibinfo {author} {\bibfnamefont {S.}~\bibnamefont {H\"ofling}}, \bibinfo
  {author} {\bibfnamefont {M.}~\bibnamefont {Kamp}}, \ and\ \bibinfo {author}
  {\bibfnamefont {L.}~\bibnamefont {Worschech}},\ }\bibfield  {title} {\enquote
  {\bibinfo {title} {Voltage fluctuation to current converter with
  coulomb-coupled quantum dots},}\ }\href {\doibase
  10.1103/PhysRevLett.114.146805} {\bibfield  {journal} {\bibinfo  {journal}
  {Phys. Rev. Lett.}\ }\textbf {\bibinfo {volume} {114}},\ \bibinfo {pages}
  {146805} (\bibinfo {year} {2015})}\BibitemShut {NoStop}%
\bibitem [{\citenamefont {Thierschmann}\ \emph {et~al.}(2015)\citenamefont
  {Thierschmann}, \citenamefont {S{\'a}nchez}, \citenamefont {Sothmann},
  \citenamefont {Arnold}, \citenamefont {Heyn}, \citenamefont {Hansen},
  \citenamefont {Buhmann},\ and\ \citenamefont {Molenkamp}}]{Thier2015}%
  \BibitemOpen
  \bibfield  {author} {\bibinfo {author} {\bibfnamefont {H.}~\bibnamefont
  {Thierschmann}}, \bibinfo {author} {\bibfnamefont {R.}~\bibnamefont
  {S{\'a}nchez}}, \bibinfo {author} {\bibfnamefont {B.}~\bibnamefont
  {Sothmann}}, \bibinfo {author} {\bibfnamefont {F.}~\bibnamefont {Arnold}},
  \bibinfo {author} {\bibfnamefont {C.}~\bibnamefont {Heyn}}, \bibinfo {author}
  {\bibfnamefont {W.}~\bibnamefont {Hansen}}, \bibinfo {author} {\bibfnamefont
  {H.}~\bibnamefont {Buhmann}}, \ and\ \bibinfo {author} {\bibfnamefont
  {L.~W.}\ \bibnamefont {Molenkamp}},\ }\bibfield  {title} {\enquote {\bibinfo
  {title} {Three-terminal energy harvester with coupled quantum dots},}\ }\href
  {\doibase 10.1038/nnano.2015.176} {\bibfield  {journal} {\bibinfo  {journal}
  {Nat. Nanotech.}\ }\textbf {\bibinfo {volume} {10}},\ \bibinfo {pages} {854}
  (\bibinfo {year} {2015})}\BibitemShut {NoStop}%
\bibitem [{\citenamefont {Mart{\'\i}nez}\ \emph {et~al.}(2016)\citenamefont
  {Mart{\'\i}nez}, \citenamefont {Rold{\'a}n}, \citenamefont {Dinis},
  \citenamefont {Petrov}, \citenamefont {Parrondo},\ and\ \citenamefont
  {Rica}}]{NP}%
  \BibitemOpen
  \bibfield  {author} {\bibinfo {author} {\bibfnamefont {I.~A.}\ \bibnamefont
  {Mart{\'\i}nez}}, \bibinfo {author} {\bibfnamefont {{\'E}.}~\bibnamefont
  {Rold{\'a}n}}, \bibinfo {author} {\bibfnamefont {L.}~\bibnamefont {Dinis}},
  \bibinfo {author} {\bibfnamefont {D.}~\bibnamefont {Petrov}}, \bibinfo
  {author} {\bibfnamefont {J.~M.~R.}\ \bibnamefont {Parrondo}}, \ and\ \bibinfo
  {author} {\bibfnamefont {R.~A}\ \bibnamefont {Rica}},\ }\bibfield  {title}
  {\enquote {\bibinfo {title} {Brownian carnot engine},}\ }\href {\doibase
  10.1038/nphys3518} {\bibfield  {journal} {\bibinfo  {journal} {Nat. Phys.}\
  }\textbf {\bibinfo {volume} {12}},\ \bibinfo {pages} {67} (\bibinfo {year}
  {2016})}\BibitemShut {NoStop}%
\bibitem [{\citenamefont {Jaliel}\ \emph {et~al.}(2019)\citenamefont {Jaliel},
  \citenamefont {Puddy}, \citenamefont {S\'anchez}, \citenamefont {Jordan},
  \citenamefont {Sothmann}, \citenamefont {Farrer}, \citenamefont {Griffiths},
  \citenamefont {Ritchie},\ and\ \citenamefont {Smith}}]{jaliel-exper}%
  \BibitemOpen
  \bibfield  {author} {\bibinfo {author} {\bibfnamefont {G.}~\bibnamefont
  {Jaliel}}, \bibinfo {author} {\bibfnamefont {R.~K.}\ \bibnamefont {Puddy}},
  \bibinfo {author} {\bibfnamefont {R.}~\bibnamefont {S\'anchez}}, \bibinfo
  {author} {\bibfnamefont {A.~N.}\ \bibnamefont {Jordan}}, \bibinfo {author}
  {\bibfnamefont {B.}~\bibnamefont {Sothmann}}, \bibinfo {author}
  {\bibfnamefont {I.}~\bibnamefont {Farrer}}, \bibinfo {author} {\bibfnamefont
  {J.~P.}\ \bibnamefont {Griffiths}}, \bibinfo {author} {\bibfnamefont {D.~A.}\
  \bibnamefont {Ritchie}}, \ and\ \bibinfo {author} {\bibfnamefont {C.~G.}\
  \bibnamefont {Smith}},\ }\bibfield  {title} {\enquote {\bibinfo {title}
  {Experimental realization of a quantum dot energy harvester},}\ }\href
  {\doibase 10.1103/PhysRevLett.123.117701} {\bibfield  {journal} {\bibinfo
  {journal} {Phys. Rev. Lett.}\ }\textbf {\bibinfo {volume} {123}},\ \bibinfo
  {pages} {117701} (\bibinfo {year} {2019})}\BibitemShut {NoStop}%
\bibitem [{\citenamefont {Li}\ \emph {et~al.}(2012)\citenamefont {Li},
  \citenamefont {Ren}, \citenamefont {Wang}, \citenamefont {Zhang},
  \citenamefont {H\"anggi},\ and\ \citenamefont {Li}}]{RenRMP}%
  \BibitemOpen
  \bibfield  {author} {\bibinfo {author} {\bibfnamefont {N.}~\bibnamefont
  {Li}}, \bibinfo {author} {\bibfnamefont {J.}~\bibnamefont {Ren}}, \bibinfo
  {author} {\bibfnamefont {L.}~\bibnamefont {Wang}}, \bibinfo {author}
  {\bibfnamefont {G.}~\bibnamefont {Zhang}}, \bibinfo {author} {\bibfnamefont
  {P.}~\bibnamefont {H\"anggi}}, \ and\ \bibinfo {author} {\bibfnamefont
  {B.}~\bibnamefont {Li}},\ }\bibfield  {title} {\enquote {\bibinfo {title}
  {Phononics: Manipulating heat flow with electronic analogs and beyond},}\
  }\href {\doibase 10.1103/RevModPhys.84.1045} {\bibfield  {journal} {\bibinfo
  {journal} {Rev. Mod. Phys.}\ }\textbf {\bibinfo {volume} {84}},\ \bibinfo
  {pages} {1045} (\bibinfo {year} {2012})}\BibitemShut {NoStop}%
\bibitem [{\citenamefont {Segal}\ and\ \citenamefont
  {Agarwalla}(2016)}]{segal2016arpc}%
  \BibitemOpen
  \bibfield  {author} {\bibinfo {author} {\bibfnamefont {D.}~\bibnamefont
  {Segal}}\ and\ \bibinfo {author} {\bibfnamefont {B.~K.}\ \bibnamefont
  {Agarwalla}},\ }\bibfield  {title} {\enquote {\bibinfo {title} {Vibrational
  heat transport in molecular junctions},}\ }\href {\doibase
  10.1146/annurev-physchem-040215-112103} {\bibfield  {journal} {\bibinfo
  {journal} {Annu. Rev. Phys. Chem.}\ }\textbf {\bibinfo {volume} {67}},\
  \bibinfo {pages} {185} (\bibinfo {year} {2016})}\BibitemShut {NoStop}%
\bibitem [{\citenamefont {Cui}\ \emph {et~al.}(2017)\citenamefont {Cui},
  \citenamefont {Jeong}, \citenamefont {Hur}, \citenamefont {Matt},
  \citenamefont {Kl{\"o}ckner}, \citenamefont {Pauly}, \citenamefont {Nielaba},
  \citenamefont {Cuevas}, \citenamefont {Meyhofer},\ and\ \citenamefont
  {Reddy}}]{Cuieaam6622}%
  \BibitemOpen
  \bibfield  {author} {\bibinfo {author} {\bibfnamefont {L.}~\bibnamefont
  {Cui}}, \bibinfo {author} {\bibfnamefont {W.}~\bibnamefont {Jeong}}, \bibinfo
  {author} {\bibfnamefont {S.}~\bibnamefont {Hur}}, \bibinfo {author}
  {\bibfnamefont {M.}~\bibnamefont {Matt}}, \bibinfo {author} {\bibfnamefont
  {J.~C.}\ \bibnamefont {Kl{\"o}ckner}}, \bibinfo {author} {\bibfnamefont
  {F.}~\bibnamefont {Pauly}}, \bibinfo {author} {\bibfnamefont
  {P.}~\bibnamefont {Nielaba}}, \bibinfo {author} {\bibfnamefont {J.~C.}\
  \bibnamefont {Cuevas}}, \bibinfo {author} {\bibfnamefont {E.}~\bibnamefont
  {Meyhofer}}, \ and\ \bibinfo {author} {\bibfnamefont {P.}~\bibnamefont
  {Reddy}},\ }\bibfield  {title} {\enquote {\bibinfo {title} {Quantized thermal
  transport in single-atom junctions},}\ }\href {\doibase
  10.1126/science.aam6622} {\bibfield  {journal} {\bibinfo  {journal}
  {Science}\ }\textbf {\bibinfo {volume} {355}},\ \bibinfo {pages} {1192}
  (\bibinfo {year} {2017})}\BibitemShut {NoStop}%
\bibitem [{\citenamefont {Zhang}\ \emph {et~al.}(2020)\citenamefont {Zhang},
  \citenamefont {Ouyang}, \citenamefont {Cheng}, \citenamefont {Chen},
  \citenamefont {Li},\ and\ \citenamefont {Zhang}}]{ZHANG20201}%
  \BibitemOpen
  \bibfield  {author} {\bibinfo {author} {\bibfnamefont {Z.}~\bibnamefont
  {Zhang}}, \bibinfo {author} {\bibfnamefont {Y.}~\bibnamefont {Ouyang}},
  \bibinfo {author} {\bibfnamefont {Y.}~\bibnamefont {Cheng}}, \bibinfo
  {author} {\bibfnamefont {J.}~\bibnamefont {Chen}}, \bibinfo {author}
  {\bibfnamefont {N.}~\bibnamefont {Li}}, \ and\ \bibinfo {author}
  {\bibfnamefont {G.}~\bibnamefont {Zhang}},\ }\bibfield  {title} {\enquote
  {\bibinfo {title} {Size-dependent phononic thermal transport in
  low-dimensional nanomaterials},}\ }\href {\doibase
  https://doi.org/10.1016/j.physrep.2020.03.001} {\bibfield  {journal}
  {\bibinfo  {journal} {Phys. Rep.}\ }\textbf {\bibinfo {volume} {860}},\
  \bibinfo {pages} {1} (\bibinfo {year} {2020})}\BibitemShut {NoStop}%
\bibitem [{\citenamefont {Li}\ \emph {et~al.}(2004)\citenamefont {Li},
  \citenamefont {Wang},\ and\ \citenamefont {Casati}}]{bli2004prl}%
  \BibitemOpen
  \bibfield  {author} {\bibinfo {author} {\bibfnamefont {B.}~\bibnamefont
  {Li}}, \bibinfo {author} {\bibfnamefont {L.}~\bibnamefont {Wang}}, \ and\
  \bibinfo {author} {\bibfnamefont {G.}~\bibnamefont {Casati}},\ }\bibfield
  {title} {\enquote {\bibinfo {title} {Thermal diode: Rectification of heat
  flux},}\ }\href {\doibase 10.1103/PhysRevLett.93.184301} {\bibfield
  {journal} {\bibinfo  {journal} {Phys. Rev. Lett.}\ }\textbf {\bibinfo
  {volume} {93}},\ \bibinfo {pages} {184301} (\bibinfo {year}
  {2004})}\BibitemShut {NoStop}%
\bibitem [{\citenamefont {Li}\ \emph {et~al.}(2006)\citenamefont {Li},
  \citenamefont {Wang},\ and\ \citenamefont {Casati}}]{bli2006apl}%
  \BibitemOpen
  \bibfield  {author} {\bibinfo {author} {\bibfnamefont {B.}~\bibnamefont
  {Li}}, \bibinfo {author} {\bibfnamefont {L.}~\bibnamefont {Wang}}, \ and\
  \bibinfo {author} {\bibfnamefont {G.}~\bibnamefont {Casati}},\ }\bibfield
  {title} {\enquote {\bibinfo {title} {Negative differential thermal resistance
  and thermal transistor},}\ }\href {\doibase 10.1063/1.2191730} {\bibfield
  {journal} {\bibinfo  {journal} {Appl. Phys. Lett.}\ }\textbf {\bibinfo
  {volume} {88}},\ \bibinfo {pages} {143501} (\bibinfo {year}
  {2006})}\BibitemShut {NoStop}%
\bibitem [{\citenamefont {Jiang}\ \emph
  {et~al.}(2015{\natexlab{a}})\citenamefont {Jiang}, \citenamefont {Kulkarni},
  \citenamefont {Segal},\ and\ \citenamefont {Imry}}]{Jiangtransistors}%
  \BibitemOpen
  \bibfield  {author} {\bibinfo {author} {\bibfnamefont {J.-H.}\ \bibnamefont
  {Jiang}}, \bibinfo {author} {\bibfnamefont {M.}~\bibnamefont {Kulkarni}},
  \bibinfo {author} {\bibfnamefont {D.}~\bibnamefont {Segal}}, \ and\ \bibinfo
  {author} {\bibfnamefont {Y.}~\bibnamefont {Imry}},\ }\bibfield  {title}
  {\enquote {\bibinfo {title} {Phonon thermoelectric transistors and
  rectifiers},}\ }\href {\doibase 10.1103/PhysRevB.92.045309} {\bibfield
  {journal} {\bibinfo  {journal} {Phys. Rev. B}\ }\textbf {\bibinfo {volume}
  {92}},\ \bibinfo {pages} {045309} (\bibinfo {year}
  {2015}{\natexlab{a}})}\BibitemShut {NoStop}%
\bibitem [{\citenamefont {Joulain}\ \emph {et~al.}(2016)\citenamefont
  {Joulain}, \citenamefont {Drevillon}, \citenamefont {Ezzahri},\ and\
  \citenamefont {Ordonez-Miranda}}]{Transistor9}%
  \BibitemOpen
  \bibfield  {author} {\bibinfo {author} {\bibfnamefont {K.}~\bibnamefont
  {Joulain}}, \bibinfo {author} {\bibfnamefont {J.}~\bibnamefont {Drevillon}},
  \bibinfo {author} {\bibfnamefont {Y.}~\bibnamefont {Ezzahri}}, \ and\
  \bibinfo {author} {\bibfnamefont {J.}~\bibnamefont {Ordonez-Miranda}},\
  }\bibfield  {title} {\enquote {\bibinfo {title} {Quantum thermal
  transistor},}\ }\href {\doibase 10.1103/PhysRevLett.116.200601} {\bibfield
  {journal} {\bibinfo  {journal} {Phys. Rev. Lett.}\ }\textbf {\bibinfo
  {volume} {116}},\ \bibinfo {pages} {200601} (\bibinfo {year}
  {2016})}\BibitemShut {NoStop}%
\bibitem [{\citenamefont {S\'anchez}\ \emph {et~al.}(2017)\citenamefont
  {S\'anchez}, \citenamefont {Thierschmann},\ and\ \citenamefont
  {Molenkamp}}]{Transistor1}%
  \BibitemOpen
  \bibfield  {author} {\bibinfo {author} {\bibfnamefont {R.}~\bibnamefont
  {S\'anchez}}, \bibinfo {author} {\bibfnamefont {H.}~\bibnamefont
  {Thierschmann}}, \ and\ \bibinfo {author} {\bibfnamefont {L.~W.}\
  \bibnamefont {Molenkamp}},\ }\bibfield  {title} {\enquote {\bibinfo {title}
  {All-thermal transistor based on stochastic switching},}\ }\href {\doibase
  10.1103/PhysRevB.95.241401} {\bibfield  {journal} {\bibinfo  {journal} {Phys.
  Rev. B}\ }\textbf {\bibinfo {volume} {95}},\ \bibinfo {pages} {241401}
  (\bibinfo {year} {2017})}\BibitemShut {NoStop}%
\bibitem [{\citenamefont {Mari}\ and\ \citenamefont
  {Eisert}(2012{\natexlab{a}})}]{amari2012prl}%
  \BibitemOpen
  \bibfield  {author} {\bibinfo {author} {\bibfnamefont {A.}~\bibnamefont
  {Mari}}\ and\ \bibinfo {author} {\bibfnamefont {J.}~\bibnamefont {Eisert}},\
  }\bibfield  {title} {\enquote {\bibinfo {title} {Cooling by heating: Very hot
  thermal light can significantly cool quantum systems},}\ }\href {\doibase
  10.1103/PhysRevLett.108.120602} {\bibfield  {journal} {\bibinfo  {journal}
  {Phys. Rev. Lett.}\ }\textbf {\bibinfo {volume} {108}},\ \bibinfo {pages}
  {120602} (\bibinfo {year} {2012}{\natexlab{a}})}\BibitemShut {NoStop}%
\bibitem [{\citenamefont {Cleuren}\ \emph
  {et~al.}(2012{\natexlab{a}})\citenamefont {Cleuren}, \citenamefont {Rutten},\
  and\ \citenamefont {Van~den Broeck}}]{bcleuren2012prl}%
  \BibitemOpen
  \bibfield  {author} {\bibinfo {author} {\bibfnamefont {B.}~\bibnamefont
  {Cleuren}}, \bibinfo {author} {\bibfnamefont {B.}~\bibnamefont {Rutten}}, \
  and\ \bibinfo {author} {\bibfnamefont {C.}~\bibnamefont {Van~den Broeck}},\
  }\bibfield  {title} {\enquote {\bibinfo {title} {Cooling by heating:
  Refrigeration powered by photons},}\ }\href {\doibase
  10.1103/PhysRevLett.108.120603} {\bibfield  {journal} {\bibinfo  {journal}
  {Phys. Rev. Lett.}\ }\textbf {\bibinfo {volume} {108}},\ \bibinfo {pages}
  {120603} (\bibinfo {year} {2012}{\natexlab{a}})}\BibitemShut {NoStop}%
\bibitem [{\citenamefont {H\"artle}\ \emph
  {et~al.}(2018{\natexlab{a}})\citenamefont {H\"artle}, \citenamefont
  {Schinabeck}, \citenamefont {Kulkarni}, \citenamefont {Gelbwaser-Klimovsky},
  \citenamefont {Thoss},\ and\ \citenamefont {Peskin}}]{hartle2018prb}%
  \BibitemOpen
  \bibfield  {author} {\bibinfo {author} {\bibfnamefont {R.}~\bibnamefont
  {H\"artle}}, \bibinfo {author} {\bibfnamefont {C.}~\bibnamefont
  {Schinabeck}}, \bibinfo {author} {\bibfnamefont {M.}~\bibnamefont
  {Kulkarni}}, \bibinfo {author} {\bibfnamefont {D.}~\bibnamefont
  {Gelbwaser-Klimovsky}}, \bibinfo {author} {\bibfnamefont {M.}~\bibnamefont
  {Thoss}}, \ and\ \bibinfo {author} {\bibfnamefont {U.}~\bibnamefont
  {Peskin}},\ }\bibfield  {title} {\enquote {\bibinfo {title} {Cooling by
  heating in nonequilibrium nanosystems},}\ }\href {\doibase
  10.1103/PhysRevB.98.081404} {\bibfield  {journal} {\bibinfo  {journal} {Phys.
  Rev. B}\ }\textbf {\bibinfo {volume} {98}},\ \bibinfo {pages} {081404}
  (\bibinfo {year} {2018}{\natexlab{a}})}\BibitemShut {NoStop}%
\bibitem [{\citenamefont {Entin-Wohlman}\ \emph {et~al.}(2010)\citenamefont
  {Entin-Wohlman}, \citenamefont {Imry},\ and\ \citenamefont
  {Aharony}}]{ewohlman2010prb}%
  \BibitemOpen
  \bibfield  {author} {\bibinfo {author} {\bibfnamefont {O.}~\bibnamefont
  {Entin-Wohlman}}, \bibinfo {author} {\bibfnamefont {Y.}~\bibnamefont {Imry}},
  \ and\ \bibinfo {author} {\bibfnamefont {A.}~\bibnamefont {Aharony}},\
  }\bibfield  {title} {\enquote {\bibinfo {title} {Three-terminal
  thermoelectric transport through a molecular junction},}\ }\href {\doibase
  10.1103/PhysRevB.82.115314} {\bibfield  {journal} {\bibinfo  {journal} {Phys.
  Rev. B}\ }\textbf {\bibinfo {volume} {82}},\ \bibinfo {pages} {115314}
  (\bibinfo {year} {2010})}\BibitemShut {NoStop}%
\bibitem [{\citenamefont {S\'anchez}\ and\ \citenamefont
  {B\"uttiker}(2011{\natexlab{b}})}]{rsanchez2011prb}%
  \BibitemOpen
  \bibfield  {author} {\bibinfo {author} {\bibfnamefont {R.}~\bibnamefont
  {S\'anchez}}\ and\ \bibinfo {author} {\bibfnamefont {M.}~\bibnamefont
  {B\"uttiker}},\ }\bibfield  {title} {\enquote {\bibinfo {title} {Optimal
  energy quanta to current conversion},}\ }\href {\doibase
  10.1103/PhysRevB.83.085428} {\bibfield  {journal} {\bibinfo  {journal} {Phys.
  Rev. B}\ }\textbf {\bibinfo {volume} {83}},\ \bibinfo {pages} {085428}
  (\bibinfo {year} {2011}{\natexlab{b}})}\BibitemShut {NoStop}%
\bibitem [{\citenamefont {S\'anchez}\ and\ \citenamefont
  {Serra}(2011)}]{David2011PRB}%
  \BibitemOpen
  \bibfield  {author} {\bibinfo {author} {\bibfnamefont {D.}~\bibnamefont
  {S\'anchez}}\ and\ \bibinfo {author} {\bibfnamefont {L.}~\bibnamefont
  {Serra}},\ }\bibfield  {title} {\enquote {\bibinfo {title} {Thermoelectric
  transport of mesoscopic conductors coupled to voltage and thermal probes},}\
  }\href {\doibase 10.1103/PhysRevB.84.201307} {\bibfield  {journal} {\bibinfo
  {journal} {Phys. Rev. B}\ }\textbf {\bibinfo {volume} {84}},\ \bibinfo
  {pages} {201307} (\bibinfo {year} {2011})}\BibitemShut {NoStop}%
\bibitem [{\citenamefont {Jiang}\ \emph {et~al.}(2013)\citenamefont {Jiang},
  \citenamefont {Entin-Wohlman},\ and\ \citenamefont {Imry}}]{Jiang2013}%
  \BibitemOpen
  \bibfield  {author} {\bibinfo {author} {\bibfnamefont {J.-H.}\ \bibnamefont
  {Jiang}}, \bibinfo {author} {\bibfnamefont {O.}~\bibnamefont
  {Entin-Wohlman}}, \ and\ \bibinfo {author} {\bibfnamefont {Y.}~\bibnamefont
  {Imry}},\ }\bibfield  {title} {\enquote {\bibinfo {title} {Hopping
  thermoelectric transport in finite systems: Boundary effects},}\ }\href
  {\doibase 10.1103/PhysRevB.87.205420} {\bibfield  {journal} {\bibinfo
  {journal} {Phys. Rev. B}\ }\textbf {\bibinfo {volume} {87}},\ \bibinfo
  {pages} {205420} (\bibinfo {year} {2013})}\BibitemShut {NoStop}%
\bibitem [{\citenamefont {Verley}\ \emph
  {et~al.}(2014{\natexlab{a}})\citenamefont {Verley}, \citenamefont {Esposito},
  \citenamefont {Willaert},\ and\ \citenamefont {Van Den~Broeck}}]{Verley2014}%
  \BibitemOpen
  \bibfield  {author} {\bibinfo {author} {\bibfnamefont {G.}~\bibnamefont
  {Verley}}, \bibinfo {author} {\bibfnamefont {Massimiliano}\ \bibnamefont
  {Esposito}}, \bibinfo {author} {\bibfnamefont {T.}~\bibnamefont {Willaert}},
  \ and\ \bibinfo {author} {\bibfnamefont {C.}~\bibnamefont {Van Den~Broeck}},\
  }\bibfield  {title} {\enquote {\bibinfo {title} {The unlikely carnot
  efficiency},}\ }\href {\doibase 10.1038/ncomms5721} {\bibfield  {journal}
  {\bibinfo  {journal} {Nat. Commun.}\ }\textbf {\bibinfo {volume} {5}},\
  \bibinfo {pages} {4721} (\bibinfo {year} {2014}{\natexlab{a}})}\BibitemShut
  {NoStop}%
\bibitem [{\citenamefont {Jiang}\ and\ \citenamefont
  {Imry}(2018)}]{JiangNearfield}%
  \BibitemOpen
  \bibfield  {author} {\bibinfo {author} {\bibfnamefont {J.-H.}\ \bibnamefont
  {Jiang}}\ and\ \bibinfo {author} {\bibfnamefont {Y.}~\bibnamefont {Imry}},\
  }\bibfield  {title} {\enquote {\bibinfo {title} {Near-field three-terminal
  thermoelectric heat engine},}\ }\href {\doibase 10.1103/PhysRevB.97.125422}
  {\bibfield  {journal} {\bibinfo  {journal} {Phys. Rev. B}\ }\textbf {\bibinfo
  {volume} {97}},\ \bibinfo {pages} {125422} (\bibinfo {year}
  {2018})}\BibitemShut {NoStop}%
\bibitem [{\citenamefont {Lu}\ \emph {et~al.}(2019{\natexlab{a}})\citenamefont
  {Lu}, \citenamefont {Liu}, \citenamefont {Wang}, \citenamefont {Wang},\ and\
  \citenamefont {Jiang}}]{trade-off}%
  \BibitemOpen
  \bibfield  {author} {\bibinfo {author} {\bibfnamefont {J.}~\bibnamefont
  {Lu}}, \bibinfo {author} {\bibfnamefont {Y.}~\bibnamefont {Liu}}, \bibinfo
  {author} {\bibfnamefont {R.}~\bibnamefont {Wang}}, \bibinfo {author}
  {\bibfnamefont {C.}~\bibnamefont {Wang}}, \ and\ \bibinfo {author}
  {\bibfnamefont {J.-H.}\ \bibnamefont {Jiang}},\ }\bibfield  {title} {\enquote
  {\bibinfo {title} {Optimal efficiency and power, and their trade-off in
  three-terminal quantum thermoelectric engines with two output electric
  currents},}\ }\href {\doibase 10.1103/PhysRevB.100.115438} {\bibfield
  {journal} {\bibinfo  {journal} {Phys. Rev. B}\ }\textbf {\bibinfo {volume}
  {100}},\ \bibinfo {pages} {115438} (\bibinfo {year}
  {2019}{\natexlab{a}})}\BibitemShut {NoStop}%
\bibitem [{\citenamefont {Rowe}(2018)}]{rowe2018thermoelectrics}%
  \BibitemOpen
  \bibfield  {author} {\bibinfo {author} {\bibfnamefont {D.~M.}\ \bibnamefont
  {Rowe}},\ }\href {https://books.google.co.jp/books?id=0iwERQe5IKQC} {\emph
  {\bibinfo {title} {Thermoelectrics Handbook: Macro to Nano}}}\ (\bibinfo
  {publisher} {CRC Press},\ \bibinfo {year} {2018})\BibitemShut {NoStop}%
\bibitem [{\citenamefont {DiSalvo}(1999)}]{DiSalvo703}%
  \BibitemOpen
  \bibfield  {author} {\bibinfo {author} {\bibfnamefont {F.~J.}\ \bibnamefont
  {DiSalvo}},\ }\bibfield  {title} {\enquote {\bibinfo {title} {Thermoelectric
  cooling and power generation},}\ }\href {\doibase
  10.1126/science.285.5428.703} {\bibfield  {journal} {\bibinfo  {journal}
  {Science}\ }\textbf {\bibinfo {volume} {285}},\ \bibinfo {pages} {703}
  (\bibinfo {year} {1999})}\BibitemShut {NoStop}%
\bibitem [{\citenamefont {Whitney}(2014)}]{rwhitney2014prl}%
  \BibitemOpen
  \bibfield  {author} {\bibinfo {author} {\bibfnamefont {R.~S.}\ \bibnamefont
  {Whitney}},\ }\bibfield  {title} {\enquote {\bibinfo {title} {Most efficient
  quantum thermoelectric at finite power output},}\ }\href {\doibase
  10.1103/PhysRevLett.112.130601} {\bibfield  {journal} {\bibinfo  {journal}
  {Phys. Rev. Lett.}\ }\textbf {\bibinfo {volume} {112}},\ \bibinfo {pages}
  {130601} (\bibinfo {year} {2014})}\BibitemShut {NoStop}%
\bibitem [{\citenamefont {Entin-Wohlman}\ \emph {et~al.}(2014)\citenamefont
  {Entin-Wohlman}, \citenamefont {Jiang},\ and\ \citenamefont
  {Imry}}]{ewohlman2014pre}%
  \BibitemOpen
  \bibfield  {author} {\bibinfo {author} {\bibfnamefont {O.}~\bibnamefont
  {Entin-Wohlman}}, \bibinfo {author} {\bibfnamefont {J.-H.}\ \bibnamefont
  {Jiang}}, \ and\ \bibinfo {author} {\bibfnamefont {Y.}~\bibnamefont {Imry}},\
  }\bibfield  {title} {\enquote {\bibinfo {title} {Efficiency and dissipation
  in a two-terminal thermoelectric junction, emphasizing small dissipation},}\
  }\href {\doibase 10.1103/PhysRevE.89.012123} {\bibfield  {journal} {\bibinfo
  {journal} {Phys. Rev. E}\ }\textbf {\bibinfo {volume} {89}},\ \bibinfo
  {pages} {012123} (\bibinfo {year} {2014})}\BibitemShut {NoStop}%
\bibitem [{\citenamefont {Lu}\ \emph {et~al.}(2019{\natexlab{b}})\citenamefont
  {Lu}, \citenamefont {Zhuo}, \citenamefont {Sun},\ and\ \citenamefont
  {Jiang}}]{CooperativeSpin}%
  \BibitemOpen
  \bibfield  {author} {\bibinfo {author} {\bibfnamefont {J.~C.}\ \bibnamefont
  {Lu}}, \bibinfo {author} {\bibfnamefont {F.~J.}\ \bibnamefont {Zhuo}},
  \bibinfo {author} {\bibfnamefont {Z.~Z.}\ \bibnamefont {Sun}}, \ and\
  \bibinfo {author} {\bibfnamefont {J.~H.}\ \bibnamefont {Jiang}},\ }\bibfield
  {title} {\enquote {\bibinfo {title} {Cooperative spin caloritronic
  devices},}\ }\href {\doibase 10.30919/esee8c353} {\bibfield  {journal}
  {\bibinfo  {journal} {ES Energy. Environ.}\ }\textbf {\bibinfo {volume}
  {7}},\ \bibinfo {pages} {17} (\bibinfo {year}
  {2019}{\natexlab{b}})}\BibitemShut {NoStop}%
\bibitem [{\citenamefont {Jiang}\ \emph {et~al.}(2012)\citenamefont {Jiang},
  \citenamefont {Entin-Wohlman},\ and\ \citenamefont {Imry}}]{Jiang2012}%
  \BibitemOpen
  \bibfield  {author} {\bibinfo {author} {\bibfnamefont {J.-H.}\ \bibnamefont
  {Jiang}}, \bibinfo {author} {\bibfnamefont {O.}~\bibnamefont
  {Entin-Wohlman}}, \ and\ \bibinfo {author} {\bibfnamefont {Y.}~\bibnamefont
  {Imry}},\ }\bibfield  {title} {\enquote {\bibinfo {title} {Thermoelectric
  three-terminal hopping transport through one-dimensional nanosystems},}\
  }\href {\doibase 10.1103/PhysRevB.85.075412} {\bibfield  {journal} {\bibinfo
  {journal} {Phys. Rev. B}\ }\textbf {\bibinfo {volume} {85}},\ \bibinfo
  {pages} {075412} (\bibinfo {year} {2012})}\BibitemShut {NoStop}%
\bibitem [{\citenamefont {Sothmann}\ \emph {et~al.}(2012)\citenamefont
  {Sothmann}, \citenamefont {S\'anchez}, \citenamefont {Jordan},\ and\
  \citenamefont {B\"uttiker}}]{sothmann2012prb}%
  \BibitemOpen
  \bibfield  {author} {\bibinfo {author} {\bibfnamefont {B.}~\bibnamefont
  {Sothmann}}, \bibinfo {author} {\bibfnamefont {R.}~\bibnamefont {S\'anchez}},
  \bibinfo {author} {\bibfnamefont {A.~N.}\ \bibnamefont {Jordan}}, \ and\
  \bibinfo {author} {\bibfnamefont {M.}~\bibnamefont {B\"uttiker}},\ }\bibfield
   {title} {\enquote {\bibinfo {title} {Rectification of thermal fluctuations
  in a chaotic cavity heat engine},}\ }\href {\doibase
  10.1103/PhysRevB.85.205301} {\bibfield  {journal} {\bibinfo  {journal} {Phys.
  Rev. B}\ }\textbf {\bibinfo {volume} {85}},\ \bibinfo {pages} {205301}
  (\bibinfo {year} {2012})}\BibitemShut {NoStop}%
\bibitem [{\citenamefont {Lu}\ \emph {et~al.}(2019{\natexlab{c}})\citenamefont
  {Lu}, \citenamefont {Wang}, \citenamefont {Ren}, \citenamefont {Kulkarni},\
  and\ \citenamefont {Jiang}}]{lu-PRB}%
  \BibitemOpen
  \bibfield  {author} {\bibinfo {author} {\bibfnamefont {J.}~\bibnamefont
  {Lu}}, \bibinfo {author} {\bibfnamefont {R.}~\bibnamefont {Wang}}, \bibinfo
  {author} {\bibfnamefont {J.}~\bibnamefont {Ren}}, \bibinfo {author}
  {\bibfnamefont {M.}~\bibnamefont {Kulkarni}}, \ and\ \bibinfo {author}
  {\bibfnamefont {J.-H.}\ \bibnamefont {Jiang}},\ }\bibfield  {title} {\enquote
  {\bibinfo {title} {Quantum-dot circuit-qed thermoelectric diodes and
  transistors},}\ }\href {\doibase 10.1103/PhysRevB.99.035129} {\bibfield
  {journal} {\bibinfo  {journal} {Phys. Rev. B}\ }\textbf {\bibinfo {volume}
  {99}},\ \bibinfo {pages} {035129} (\bibinfo {year}
  {2019}{\natexlab{c}})}\BibitemShut {NoStop}%
\bibitem [{\citenamefont {Sivan}\ and\ \citenamefont {Imry}(1986)}]{Sivan}%
  \BibitemOpen
  \bibfield  {author} {\bibinfo {author} {\bibfnamefont {U.}~\bibnamefont
  {Sivan}}\ and\ \bibinfo {author} {\bibfnamefont {Y.}~\bibnamefont {Imry}},\
  }\bibfield  {title} {\enquote {\bibinfo {title} {Multichannel landauer
  formula for thermoelectric transport with application to thermopower near the
  mobility edge},}\ }\href {\doibase 10.1103/PhysRevB.33.551} {\bibfield
  {journal} {\bibinfo  {journal} {Phys. Rev. B}\ }\textbf {\bibinfo {volume}
  {33}},\ \bibinfo {pages} {551} (\bibinfo {year} {1986})}\BibitemShut
  {NoStop}%
\bibitem [{\citenamefont {Butcher}(1990)}]{butcher1990}%
  \BibitemOpen
  \bibfield  {author} {\bibinfo {author} {\bibfnamefont {P.~N.}\ \bibnamefont
  {Butcher}},\ }\bibfield  {title} {\enquote {\bibinfo {title} {Thermal and
  electrical transport formalism for electronic microstructures with many
  terminals},}\ }\href {http://stacks.iop.org/0953-8984/2/i=22/a=008}
  {\bibfield  {journal} {\bibinfo  {journal} {J. Phys.: Condens. Matter}\
  }\textbf {\bibinfo {volume} {2}},\ \bibinfo {pages} {4869} (\bibinfo {year}
  {1990})}\BibitemShut {NoStop}%
\bibitem [{\citenamefont {Jiang}(2014)}]{JiangPRE}%
  \BibitemOpen
  \bibfield  {author} {\bibinfo {author} {\bibfnamefont {J.-H.}\ \bibnamefont
  {Jiang}},\ }\bibfield  {title} {\enquote {\bibinfo {title} {Thermodynamic
  bounds and general properties of optimal efficiency and power in linear
  responses},}\ }\href {\doibase 10.1103/PhysRevE.90.042126} {\bibfield
  {journal} {\bibinfo  {journal} {Phys. Rev. E}\ }\textbf {\bibinfo {volume}
  {90}},\ \bibinfo {pages} {042126} (\bibinfo {year} {2014})}\BibitemShut
  {NoStop}%
\bibitem [{\citenamefont {Lu}\ \emph {et~al.}(2017)\citenamefont {Lu},
  \citenamefont {Wang}, \citenamefont {Liu},\ and\ \citenamefont
  {Jiang}}]{MyJAP}%
  \BibitemOpen
  \bibfield  {author} {\bibinfo {author} {\bibfnamefont {J.}~\bibnamefont
  {Lu}}, \bibinfo {author} {\bibfnamefont {R.}~\bibnamefont {Wang}}, \bibinfo
  {author} {\bibfnamefont {Y.}~\bibnamefont {Liu}}, \ and\ \bibinfo {author}
  {\bibfnamefont {J.-H.}\ \bibnamefont {Jiang}},\ }\bibfield  {title} {\enquote
  {\bibinfo {title} {Thermoelectric cooperative effect in three-terminal
  elastic transport through a quantum dot},}\ }\href {\doibase
  10.1063/1.4995532} {\bibfield  {journal} {\bibinfo  {journal} {J. Appl.
  Phys.}\ }\textbf {\bibinfo {volume} {122}},\ \bibinfo {pages} {044301}
  (\bibinfo {year} {2017})}\BibitemShut {NoStop}%
\bibitem [{\citenamefont {Bergenfeldt}\ \emph {et~al.}(2014)\citenamefont
  {Bergenfeldt}, \citenamefont {Samuelsson}, \citenamefont {Sothmann},
  \citenamefont {Flindt},\ and\ \citenamefont {B\"uttiker}}]{transistor0}%
  \BibitemOpen
  \bibfield  {author} {\bibinfo {author} {\bibfnamefont {C.}~\bibnamefont
  {Bergenfeldt}}, \bibinfo {author} {\bibfnamefont {P.}~\bibnamefont
  {Samuelsson}}, \bibinfo {author} {\bibfnamefont {B.}~\bibnamefont
  {Sothmann}}, \bibinfo {author} {\bibfnamefont {C.}~\bibnamefont {Flindt}}, \
  and\ \bibinfo {author} {\bibfnamefont {M.}~\bibnamefont {B\"uttiker}},\
  }\bibfield  {title} {\enquote {\bibinfo {title} {Hybrid microwave-cavity heat
  engine},}\ }\href {\doibase 10.1103/PhysRevLett.112.076803} {\bibfield
  {journal} {\bibinfo  {journal} {Phys. Rev. Lett.}\ }\textbf {\bibinfo
  {volume} {112}},\ \bibinfo {pages} {076803} (\bibinfo {year}
  {2014})}\BibitemShut {NoStop}%
\bibitem [{\citenamefont {Wang}\ \emph
  {et~al.}(2018{\natexlab{a}})\citenamefont {Wang}, \citenamefont {Chen},
  \citenamefont {Sun},\ and\ \citenamefont {Ren}}]{transistor4}%
  \BibitemOpen
  \bibfield  {author} {\bibinfo {author} {\bibfnamefont {C.}~\bibnamefont
  {Wang}}, \bibinfo {author} {\bibfnamefont {X.-M.}\ \bibnamefont {Chen}},
  \bibinfo {author} {\bibfnamefont {K.-W.}\ \bibnamefont {Sun}}, \ and\
  \bibinfo {author} {\bibfnamefont {J.}~\bibnamefont {Ren}},\ }\bibfield
  {title} {\enquote {\bibinfo {title} {Heat amplification and negative
  differential thermal conductance in a strongly coupled nonequilibrium
  spin-boson system},}\ }\href {\doibase 10.1103/PhysRevA.97.052112} {\bibfield
   {journal} {\bibinfo  {journal} {Phys. Rev. A}\ }\textbf {\bibinfo {volume}
  {97}},\ \bibinfo {pages} {052112} (\bibinfo {year}
  {2018}{\natexlab{a}})}\BibitemShut {NoStop}%
\bibitem [{\citenamefont {Yang}\ \emph {et~al.}(2019)\citenamefont {Yang},
  \citenamefont {Elouard}, \citenamefont {Splettstoesser}, \citenamefont
  {Sothmann}, \citenamefont {S\'anchez},\ and\ \citenamefont
  {Jordan}}]{transistor-yang}%
  \BibitemOpen
  \bibfield  {author} {\bibinfo {author} {\bibfnamefont {J.}~\bibnamefont
  {Yang}}, \bibinfo {author} {\bibfnamefont {C.}~\bibnamefont {Elouard}},
  \bibinfo {author} {\bibfnamefont {J.}~\bibnamefont {Splettstoesser}},
  \bibinfo {author} {\bibfnamefont {B.}~\bibnamefont {Sothmann}}, \bibinfo
  {author} {\bibfnamefont {R.}~\bibnamefont {S\'anchez}}, \ and\ \bibinfo
  {author} {\bibfnamefont {A.~N.}\ \bibnamefont {Jordan}},\ }\bibfield  {title}
  {\enquote {\bibinfo {title} {Thermal transistor and thermometer based on
  coulomb-coupled conductors},}\ }\href {\doibase 10.1103/PhysRevB.100.045418}
  {\bibfield  {journal} {\bibinfo  {journal} {Phys. Rev. B}\ }\textbf {\bibinfo
  {volume} {100}},\ \bibinfo {pages} {045418} (\bibinfo {year}
  {2019})}\BibitemShut {NoStop}%
\bibitem [{\citenamefont {Liu}\ \emph {et~al.}(2019)\citenamefont {Liu},
  \citenamefont {Wang}, \citenamefont {Wang},\ and\ \citenamefont
  {Ren}}]{WangPRE}%
  \BibitemOpen
  \bibfield  {author} {\bibinfo {author} {\bibfnamefont {H.}~\bibnamefont
  {Liu}}, \bibinfo {author} {\bibfnamefont {C.}~\bibnamefont {Wang}}, \bibinfo
  {author} {\bibfnamefont {L.-Q.}\ \bibnamefont {Wang}}, \ and\ \bibinfo
  {author} {\bibfnamefont {J.}~\bibnamefont {Ren}},\ }\bibfield  {title}
  {\enquote {\bibinfo {title} {Strong system-bath coupling induces negative
  differential thermal conductance and heat amplification in nonequilibrium
  two-qubit systems},}\ }\href {\doibase 10.1103/PhysRevE.99.032114} {\bibfield
   {journal} {\bibinfo  {journal} {Phys. Rev. E}\ }\textbf {\bibinfo {volume}
  {99}},\ \bibinfo {pages} {032114} (\bibinfo {year} {2019})}\BibitemShut
  {NoStop}%
\bibitem [{\citenamefont {Mari}\ and\ \citenamefont
  {Eisert}(2012{\natexlab{b}})}]{Cooling1}%
  \BibitemOpen
  \bibfield  {author} {\bibinfo {author} {\bibfnamefont {A.}~\bibnamefont
  {Mari}}\ and\ \bibinfo {author} {\bibfnamefont {J.}~\bibnamefont {Eisert}},\
  }\bibfield  {title} {\enquote {\bibinfo {title} {Cooling by heating: Very hot
  thermal light can significantly cool quantum systems},}\ }\href {\doibase
  10.1103/PhysRevLett.108.120602} {\bibfield  {journal} {\bibinfo  {journal}
  {Phys. Rev. Lett.}\ }\textbf {\bibinfo {volume} {108}},\ \bibinfo {pages}
  {120602} (\bibinfo {year} {2012}{\natexlab{b}})}\BibitemShut {NoStop}%
\bibitem [{\citenamefont {Cleuren}\ \emph
  {et~al.}(2012{\natexlab{b}})\citenamefont {Cleuren}, \citenamefont {Rutten},\
  and\ \citenamefont {Van~den Broeck}}]{Cooling2}%
  \BibitemOpen
  \bibfield  {author} {\bibinfo {author} {\bibfnamefont {B.}~\bibnamefont
  {Cleuren}}, \bibinfo {author} {\bibfnamefont {B.}~\bibnamefont {Rutten}}, \
  and\ \bibinfo {author} {\bibfnamefont {C.}~\bibnamefont {Van~den Broeck}},\
  }\bibfield  {title} {\enquote {\bibinfo {title} {Cooling by heating:
  Refrigeration powered by photons},}\ }\href {\doibase
  10.1103/PhysRevLett.108.120603} {\bibfield  {journal} {\bibinfo  {journal}
  {Phys. Rev. Lett.}\ }\textbf {\bibinfo {volume} {108}},\ \bibinfo {pages}
  {120603} (\bibinfo {year} {2012}{\natexlab{b}})}\BibitemShut {NoStop}%
\bibitem [{\citenamefont {H\"artle}\ \emph
  {et~al.}(2018{\natexlab{b}})\citenamefont {H\"artle}, \citenamefont
  {Schinabeck}, \citenamefont {Kulkarni}, \citenamefont {Gelbwaser-Klimovsky},
  \citenamefont {Thoss},\ and\ \citenamefont {Peskin}}]{Cooling3}%
  \BibitemOpen
  \bibfield  {author} {\bibinfo {author} {\bibfnamefont {R.}~\bibnamefont
  {H\"artle}}, \bibinfo {author} {\bibfnamefont {C.}~\bibnamefont
  {Schinabeck}}, \bibinfo {author} {\bibfnamefont {M.}~\bibnamefont
  {Kulkarni}}, \bibinfo {author} {\bibfnamefont {D.}~\bibnamefont
  {Gelbwaser-Klimovsky}}, \bibinfo {author} {\bibfnamefont {M.}~\bibnamefont
  {Thoss}}, \ and\ \bibinfo {author} {\bibfnamefont {U.}~\bibnamefont
  {Peskin}},\ }\bibfield  {title} {\enquote {\bibinfo {title} {Cooling by
  heating in nonequilibrium nanosystems},}\ }\href {\doibase
  10.1103/PhysRevB.98.081404} {\bibfield  {journal} {\bibinfo  {journal} {Phys.
  Rev. B}\ }\textbf {\bibinfo {volume} {98}},\ \bibinfo {pages} {081404}
  (\bibinfo {year} {2018}{\natexlab{b}})}\BibitemShut {NoStop}%
\bibitem [{\citenamefont {Micadei}\ \emph {et~al.}(2019)\citenamefont
  {Micadei}, \citenamefont {Peterson}, \citenamefont {Souza}, \citenamefont
  {Sarthour}, \citenamefont {Oliveira}, \citenamefont {Landi}, \citenamefont
  {Batalh{\~a}o}, \citenamefont {Serra},\ and\ \citenamefont
  {Lutz}}]{cooling7}%
  \BibitemOpen
  \bibfield  {author} {\bibinfo {author} {\bibfnamefont {K.}~\bibnamefont
  {Micadei}}, \bibinfo {author} {\bibfnamefont {J.~P.~S.}\ \bibnamefont
  {Peterson}}, \bibinfo {author} {\bibfnamefont {A.~M.}\ \bibnamefont {Souza}},
  \bibinfo {author} {\bibfnamefont {R.~S.}\ \bibnamefont {Sarthour}}, \bibinfo
  {author} {\bibfnamefont {I.~S.}\ \bibnamefont {Oliveira}}, \bibinfo {author}
  {\bibfnamefont {G.~T.}\ \bibnamefont {Landi}}, \bibinfo {author}
  {\bibfnamefont {T.~B.}\ \bibnamefont {Batalh{\~a}o}}, \bibinfo {author}
  {\bibfnamefont {R.~M.}\ \bibnamefont {Serra}}, \ and\ \bibinfo {author}
  {\bibfnamefont {E.}~\bibnamefont {Lutz}},\ }\bibfield  {title} {\enquote
  {\bibinfo {title} {Reversing the direction of heat flow using quantum
  correlations},}\ }\href {\doibase https://doi.org/10.1038/s41467-019-10333-7}
  {\bibfield  {journal} {\bibinfo  {journal} {Nat. Commun.}\ }\textbf {\bibinfo
  {volume} {10}},\ \bibinfo {pages} {2456} (\bibinfo {year}
  {2019})}\BibitemShut {NoStop}%
\bibitem [{\citenamefont {Seifert}(2012)}]{SeifertPR}%
  \BibitemOpen
  \bibfield  {author} {\bibinfo {author} {\bibfnamefont {U.}~\bibnamefont
  {Seifert}},\ }\bibfield  {title} {\enquote {\bibinfo {title} {Stochastic
  thermodynamics, fluctuation theorems and molecular machines},}\ }\href
  {\doibase 10.1088/0034-4885/75/12/126001} {\bibfield  {journal} {\bibinfo
  {journal} {Rep. Prog. Phys.}\ }\textbf {\bibinfo {volume} {75}},\ \bibinfo
  {pages} {126001} (\bibinfo {year} {2012})}\BibitemShut {NoStop}%
\bibitem [{\citenamefont {Andrieux}\ and\ \citenamefont
  {Gaspard}(2004{\natexlab{a}})}]{PDF1}%
  \BibitemOpen
  \bibfield  {author} {\bibinfo {author} {\bibfnamefont {D.}~\bibnamefont
  {Andrieux}}\ and\ \bibinfo {author} {\bibfnamefont {P.}~\bibnamefont
  {Gaspard}},\ }\bibfield  {title} {\enquote {\bibinfo {title} {Fluctuation
  theorem and onsager reciprocity relations},}\ }\href {\doibase
  10.1063/1.1782391} {\bibfield  {journal} {\bibinfo  {journal} {J. Chem.
  Phys.}\ }\textbf {\bibinfo {volume} {121}},\ \bibinfo {pages} {6167}
  (\bibinfo {year} {2004}{\natexlab{a}})}\BibitemShut {NoStop}%
\bibitem [{\citenamefont {Gaspard}(2013{\natexlab{a}})}]{PDF2}%
  \BibitemOpen
  \bibfield  {author} {\bibinfo {author} {\bibfnamefont {P.}~\bibnamefont
  {Gaspard}},\ }\bibfield  {title} {\enquote {\bibinfo {title} {Multivariate
  fluctuation relations for currents},}\ }\href {\doibase
  10.1088/1367-2630/15/11/115014} {\bibfield  {journal} {\bibinfo  {journal}
  {New J. Phys.}\ }\textbf {\bibinfo {volume} {15}},\ \bibinfo {pages} {115014}
  (\bibinfo {year} {2013}{\natexlab{a}})}\BibitemShut {NoStop}%
\bibitem [{\citenamefont {Polettini}\ \emph {et~al.}(2015)\citenamefont
  {Polettini}, \citenamefont {Verley},\ and\ \citenamefont {Esposito}}]{PDF3}%
  \BibitemOpen
  \bibfield  {author} {\bibinfo {author} {\bibfnamefont {M.}~\bibnamefont
  {Polettini}}, \bibinfo {author} {\bibfnamefont {G.}~\bibnamefont {Verley}}, \
  and\ \bibinfo {author} {\bibfnamefont {M.}~\bibnamefont {Esposito}},\
  }\bibfield  {title} {\enquote {\bibinfo {title} {Efficiency statistics at all
  times: Carnot limit at finite power},}\ }\href {\doibase
  10.1103/PhysRevLett.114.050601} {\bibfield  {journal} {\bibinfo  {journal}
  {Phys. Rev. Lett.}\ }\textbf {\bibinfo {volume} {114}},\ \bibinfo {pages}
  {050601} (\bibinfo {year} {2015})}\BibitemShut {NoStop}%
\bibitem [{\citenamefont {Proesmans}\ \emph
  {et~al.}(2016{\natexlab{a}})\citenamefont {Proesmans}, \citenamefont
  {Dreher}, \citenamefont {Gavrilov}, \citenamefont {Bechhoefer},\ and\
  \citenamefont {Van~den Broeck}}]{Brownian}%
  \BibitemOpen
  \bibfield  {author} {\bibinfo {author} {\bibfnamefont {K.}~\bibnamefont
  {Proesmans}}, \bibinfo {author} {\bibfnamefont {Y.}~\bibnamefont {Dreher}},
  \bibinfo {author} {\bibfnamefont {M.}~\bibnamefont {Gavrilov}}, \bibinfo
  {author} {\bibfnamefont {J.}~\bibnamefont {Bechhoefer}}, \ and\ \bibinfo
  {author} {\bibfnamefont {C.}~\bibnamefont {Van~den Broeck}},\ }\bibfield
  {title} {\enquote {\bibinfo {title} {Brownian duet: A novel tale of
  thermodynamic efficiency},}\ }\href {\doibase 10.1103/PhysRevX.6.041010}
  {\bibfield  {journal} {\bibinfo  {journal} {Phys. Rev. X}\ }\textbf {\bibinfo
  {volume} {6}},\ \bibinfo {pages} {041010} (\bibinfo {year}
  {2016}{\natexlab{a}})}\BibitemShut {NoStop}%
\bibitem [{\citenamefont {Jiang}\ \emph
  {et~al.}(2015{\natexlab{b}})\citenamefont {Jiang}, \citenamefont
  {Agarwalla},\ and\ \citenamefont {Segal}}]{JiangPRL}%
  \BibitemOpen
  \bibfield  {author} {\bibinfo {author} {\bibfnamefont {J.-H.}\ \bibnamefont
  {Jiang}}, \bibinfo {author} {\bibfnamefont {B.~K.}\ \bibnamefont
  {Agarwalla}}, \ and\ \bibinfo {author} {\bibfnamefont {D.}~\bibnamefont
  {Segal}},\ }\bibfield  {title} {\enquote {\bibinfo {title} {Efficiency
  statistics and bounds for systems with broken time-reversal symmetry},}\
  }\href {\doibase 10.1103/PhysRevLett.115.040601} {\bibfield  {journal}
  {\bibinfo  {journal} {Phys. Rev. Lett.}\ }\textbf {\bibinfo {volume} {115}},\
  \bibinfo {pages} {040601} (\bibinfo {year} {2015}{\natexlab{b}})}\BibitemShut
  {NoStop}%
\bibitem [{\citenamefont {Andrieux}\ and\ \citenamefont
  {Gaspard}(2004{\natexlab{b}})}]{Gaspard}%
  \BibitemOpen
  \bibfield  {author} {\bibinfo {author} {\bibfnamefont {D.}~\bibnamefont
  {Andrieux}}\ and\ \bibinfo {author} {\bibfnamefont {P.}~\bibnamefont
  {Gaspard}},\ }\bibfield  {title} {\enquote {\bibinfo {title} {Fluctuation
  theorem and onsager reciprocity relations},}\ }\href {\doibase
  10.1063/1.1782391} {\bibfield  {journal} {\bibinfo  {journal} {J. Chem.
  Phys.}\ }\textbf {\bibinfo {volume} {121}},\ \bibinfo {pages} {6167--6174}
  (\bibinfo {year} {2004}{\natexlab{b}})}\BibitemShut {NoStop}%
\bibitem [{\citenamefont {Gaspard}(2013{\natexlab{b}})}]{Gaspard_2013}%
  \BibitemOpen
  \bibfield  {author} {\bibinfo {author} {\bibfnamefont {P.}~\bibnamefont
  {Gaspard}},\ }\bibfield  {title} {\enquote {\bibinfo {title} {Multivariate
  fluctuation relations for currents},}\ }\href {\doibase
  10.1088/1367-2630/15/11/115014} {\bibfield  {journal} {\bibinfo  {journal}
  {New J. Phys.}\ }\textbf {\bibinfo {volume} {15}},\ \bibinfo {pages} {115014}
  (\bibinfo {year} {2013}{\natexlab{b}})}\BibitemShut {NoStop}%
\bibitem [{\citenamefont {Proesmans}\ \emph
  {et~al.}(2016{\natexlab{b}})\citenamefont {Proesmans}, \citenamefont
  {Cleuren},\ and\ \citenamefont {Van~den Broeck}}]{PED}%
  \BibitemOpen
  \bibfield  {author} {\bibinfo {author} {\bibfnamefont {K.}~\bibnamefont
  {Proesmans}}, \bibinfo {author} {\bibfnamefont {B.}~\bibnamefont {Cleuren}},
  \ and\ \bibinfo {author} {\bibfnamefont {C.}~\bibnamefont {Van~den Broeck}},\
  }\bibfield  {title} {\enquote {\bibinfo {title} {Power-efficiency-dissipation
  relations in linear thermodynamics},}\ }\href {\doibase
  10.1103/PhysRevLett.116.220601} {\bibfield  {journal} {\bibinfo  {journal}
  {Phys. Rev. Lett.}\ }\textbf {\bibinfo {volume} {116}},\ \bibinfo {pages}
  {220601} (\bibinfo {year} {2016}{\natexlab{b}})}\BibitemShut {NoStop}%
\bibitem [{\citenamefont {Agarwalla}\ \emph {et~al.}(2017)\citenamefont
  {Agarwalla}, \citenamefont {Jiang},\ and\ \citenamefont
  {Segal}}]{JiangBijayPRB17}%
  \BibitemOpen
  \bibfield  {author} {\bibinfo {author} {\bibfnamefont {B.~K.}\ \bibnamefont
  {Agarwalla}}, \bibinfo {author} {\bibfnamefont {J.-H.}\ \bibnamefont
  {Jiang}}, \ and\ \bibinfo {author} {\bibfnamefont {D.}~\bibnamefont
  {Segal}},\ }\bibfield  {title} {\enquote {\bibinfo {title} {Quantum
  efficiency bound for continuous heat engines coupled to noncanonical
  reservoirs},}\ }\href {\doibase 10.1103/PhysRevB.96.104304} {\bibfield
  {journal} {\bibinfo  {journal} {Phys. Rev. B}\ }\textbf {\bibinfo {volume}
  {96}},\ \bibinfo {pages} {104304} (\bibinfo {year} {2017})}\BibitemShut
  {NoStop}%
\bibitem [{\citenamefont {Wang}\ \emph
  {et~al.}(2018{\natexlab{b}})\citenamefont {Wang}, \citenamefont {Lu},
  \citenamefont {Wang},\ and\ \citenamefont {Jiang}}]{Rongqian}%
  \BibitemOpen
  \bibfield  {author} {\bibinfo {author} {\bibfnamefont {R.}~\bibnamefont
  {Wang}}, \bibinfo {author} {\bibfnamefont {J.}~\bibnamefont {Lu}}, \bibinfo
  {author} {\bibfnamefont {C.}~\bibnamefont {Wang}}, \ and\ \bibinfo {author}
  {\bibfnamefont {J.-H.}\ \bibnamefont {Jiang}},\ }\bibfield  {title} {\enquote
  {\bibinfo {title} {Nonlinear effects for three-terminal heat engine and
  refrigerator},}\ }\href {\doibase 10.1038/s41598-018-20757-8} {\bibfield
  {journal} {\bibinfo  {journal} {Sci. Rep.}\ }\textbf {\bibinfo {volume}
  {8}},\ \bibinfo {pages} {2607} (\bibinfo {year}
  {2018}{\natexlab{b}})}\BibitemShut {NoStop}%
\bibitem [{\citenamefont {S\'anchez}\ \emph {et~al.}(2019)\citenamefont
  {S\'anchez}, \citenamefont {S\'anchez}, \citenamefont {L\'opez},\ and\
  \citenamefont {Sothmann}}]{David-refrigerator}%
  \BibitemOpen
  \bibfield  {author} {\bibinfo {author} {\bibfnamefont {D.}~\bibnamefont
  {S\'anchez}}, \bibinfo {author} {\bibfnamefont {R.}~\bibnamefont
  {S\'anchez}}, \bibinfo {author} {\bibfnamefont {R.}~\bibnamefont {L\'opez}},
  \ and\ \bibinfo {author} {\bibfnamefont {B.}~\bibnamefont {Sothmann}},\
  }\bibfield  {title} {\enquote {\bibinfo {title} {Nonlinear chiral
  refrigerators},}\ }\href {\doibase 10.1103/PhysRevB.99.245304} {\bibfield
  {journal} {\bibinfo  {journal} {Phys. Rev. B}\ }\textbf {\bibinfo {volume}
  {99}},\ \bibinfo {pages} {245304} (\bibinfo {year} {2019})}\BibitemShut
  {NoStop}%
\bibitem [{\citenamefont {Friedman}\ \emph {et~al.}(2018)\citenamefont
  {Friedman}, \citenamefont {Agarwalla},\ and\ \citenamefont
  {Segal}}]{Friedman2018}%
  \BibitemOpen
  \bibfield  {author} {\bibinfo {author} {\bibfnamefont {H.~M.}\ \bibnamefont
  {Friedman}}, \bibinfo {author} {\bibfnamefont {B.~K.}\ \bibnamefont
  {Agarwalla}}, \ and\ \bibinfo {author} {\bibfnamefont {D.}~\bibnamefont
  {Segal}},\ }\bibfield  {title} {\enquote {\bibinfo {title} {Quantum energy
  exchange and refrigeration: a full-counting statistics approach},}\ }\href
  {\doibase 10.1088/1367-2630/aad5fc} {\bibfield  {journal} {\bibinfo
  {journal} {New J. Phys.}\ }\textbf {\bibinfo {volume} {20}},\ \bibinfo
  {pages} {083026} (\bibinfo {year} {2018})}\BibitemShut {NoStop}%
\bibitem [{\citenamefont {S\'anchez}\ \emph {et~al.}(2018)\citenamefont
  {S\'anchez}, \citenamefont {Burset},\ and\ \citenamefont
  {Yeyati}}]{Rafael-Cooling}%
  \BibitemOpen
  \bibfield  {author} {\bibinfo {author} {\bibfnamefont {R.}~\bibnamefont
  {S\'anchez}}, \bibinfo {author} {\bibfnamefont {P.}~\bibnamefont {Burset}}, \
  and\ \bibinfo {author} {\bibfnamefont {A.~L.}\ \bibnamefont {Yeyati}},\
  }\bibfield  {title} {\enquote {\bibinfo {title} {Cooling by cooper pair
  splitting},}\ }\href {\doibase 10.1103/PhysRevB.98.241414} {\bibfield
  {journal} {\bibinfo  {journal} {Phys. Rev. B}\ }\textbf {\bibinfo {volume}
  {98}},\ \bibinfo {pages} {241414} (\bibinfo {year} {2018})}\BibitemShut
  {NoStop}%
\bibitem [{\citenamefont {Adams}\ \emph {et~al.}(2019)\citenamefont {Adams},
  \citenamefont {Verosky}, \citenamefont {Zebarjadi},\ and\ \citenamefont
  {Heremans}}]{cooling4}%
  \BibitemOpen
  \bibfield  {author} {\bibinfo {author} {\bibfnamefont {M.J.}\ \bibnamefont
  {Adams}}, \bibinfo {author} {\bibfnamefont {M.}~\bibnamefont {Verosky}},
  \bibinfo {author} {\bibfnamefont {M.}~\bibnamefont {Zebarjadi}}, \ and\
  \bibinfo {author} {\bibfnamefont {J.P.}\ \bibnamefont {Heremans}},\
  }\bibfield  {title} {\enquote {\bibinfo {title} {Active peltier coolers based
  on correlated and magnon-drag metals},}\ }\href {\doibase
  10.1103/PhysRevApplied.11.054008} {\bibfield  {journal} {\bibinfo  {journal}
  {Phys. Rev. Applied}\ }\textbf {\bibinfo {volume} {11}},\ \bibinfo {pages}
  {054008} (\bibinfo {year} {2019})}\BibitemShut {NoStop}%
\bibitem [{\citenamefont {Verley}\ \emph
  {et~al.}(2014{\natexlab{b}})\citenamefont {Verley}, \citenamefont {Willaert},
  \citenamefont {Van~den Broeck},\ and\ \citenamefont {Esposito}}]{FT6}%
  \BibitemOpen
  \bibfield  {author} {\bibinfo {author} {\bibfnamefont {G.}~\bibnamefont
  {Verley}}, \bibinfo {author} {\bibfnamefont {T.}~\bibnamefont {Willaert}},
  \bibinfo {author} {\bibfnamefont {C.}~\bibnamefont {Van~den Broeck}}, \ and\
  \bibinfo {author} {\bibfnamefont {M.}~\bibnamefont {Esposito}},\ }\bibfield
  {title} {\enquote {\bibinfo {title} {Universal theory of efficiency
  fluctuations},}\ }\href {\doibase 10.1103/PhysRevE.90.052145} {\bibfield
  {journal} {\bibinfo  {journal} {Phys. Rev. E}\ }\textbf {\bibinfo {volume}
  {90}},\ \bibinfo {pages} {052145} (\bibinfo {year}
  {2014}{\natexlab{b}})}\BibitemShut {NoStop}%
\bibitem [{\citenamefont {Esposito}\ \emph {et~al.}(2015)\citenamefont
  {Esposito}, \citenamefont {Ochoa},\ and\ \citenamefont {Galperin}}]{FT3}%
  \BibitemOpen
  \bibfield  {author} {\bibinfo {author} {\bibfnamefont {M.}~\bibnamefont
  {Esposito}}, \bibinfo {author} {\bibfnamefont {M.~A.}\ \bibnamefont {Ochoa}},
  \ and\ \bibinfo {author} {\bibfnamefont {M.}~\bibnamefont {Galperin}},\
  }\bibfield  {title} {\enquote {\bibinfo {title} {Efficiency fluctuations in
  quantum thermoelectric devices},}\ }\href {\doibase
  10.1103/PhysRevB.91.115417} {\bibfield  {journal} {\bibinfo  {journal} {Phys.
  Rev. B}\ }\textbf {\bibinfo {volume} {91}},\ \bibinfo {pages} {115417}
  (\bibinfo {year} {2015})}\BibitemShut {NoStop}%
\bibitem [{\citenamefont {Niedenzu}\ \emph {et~al.}(2018)\citenamefont
  {Niedenzu}, \citenamefont {Mukherjee}, \citenamefont {Ghosh}, \citenamefont
  {Kofman},\ and\ \citenamefont {Kurizki}}]{Niedenzu}%
  \BibitemOpen
  \bibfield  {author} {\bibinfo {author} {\bibfnamefont {W.}~\bibnamefont
  {Niedenzu}}, \bibinfo {author} {\bibfnamefont {V.}~\bibnamefont {Mukherjee}},
  \bibinfo {author} {\bibfnamefont {A.}~\bibnamefont {Ghosh}}, \bibinfo
  {author} {\bibfnamefont {A.~G.}\ \bibnamefont {Kofman}}, \ and\ \bibinfo
  {author} {\bibfnamefont {G.}~\bibnamefont {Kurizki}},\ }\bibfield  {title}
  {\enquote {\bibinfo {title} {Quantum engine efficiency bound beyond the
  second law of thermodynamics},}\ }\href
  {https://www.nature.com/articles/s41467-017-01991-6} {\bibfield  {journal}
  {\bibinfo  {journal} {Nat. Commun.}\ }\textbf {\bibinfo {volume} {9}},\
  \bibinfo {pages} {165} (\bibinfo {year} {2018})}\BibitemShut {NoStop}%
\bibitem [{\citenamefont {Manikandan}\ \emph {et~al.}(2019)\citenamefont
  {Manikandan}, \citenamefont {Dabelow}, \citenamefont {Eichhorn},\ and\
  \citenamefont {Krishnamurthy}}]{FT8}%
  \BibitemOpen
  \bibfield  {author} {\bibinfo {author} {\bibfnamefont {S.~K.}\ \bibnamefont
  {Manikandan}}, \bibinfo {author} {\bibfnamefont {L.}~\bibnamefont {Dabelow}},
  \bibinfo {author} {\bibfnamefont {R.}~\bibnamefont {Eichhorn}}, \ and\
  \bibinfo {author} {\bibfnamefont {S.}~\bibnamefont {Krishnamurthy}},\
  }\bibfield  {title} {\enquote {\bibinfo {title} {Efficiency fluctuations in
  microscopic machines},}\ }\href {\doibase 10.1103/PhysRevLett.122.140601}
  {\bibfield  {journal} {\bibinfo  {journal} {Phys. Rev. Lett.}\ }\textbf
  {\bibinfo {volume} {122}},\ \bibinfo {pages} {140601} (\bibinfo {year}
  {2019})}\BibitemShut {NoStop}%
\end{thebibliography}%

\end{document}